\newcommand{\ignore}[1]{}
\documentclass[12pt]{article}
\usepackage{placeins}
\usepackage[left]{lineno}
\usepackage{amsmath}
\usepackage{amssymb}
\usepackage{cancel}
 \usepackage{graphicx}
\usepackage{braket}
\usepackage{authblk}
\usepackage{caption,subcaption}
\usepackage{comment}
\usepackage{enumitem}
\usepackage[utf8]{inputenc}
\usepackage[pdftex, pdfstartview={FitH}, pdfnewwindow=true, colorlinks=false, pdfpagemode=UseNone]{hyperref}
\numberwithin{equation}{section}
\numberwithin{figure}{section}
\graphicspath{{fig/}}
\renewcommand{\baselinestretch}{1.1}
  \newlength{\abstractwidth}
\usepackage[usenames,dvipsnames,svgnames,table]{xcolor}

\newcommand{\be}{\begin{equation}}
 \newcommand{\ee}{\end{equation}}
\newcommand{\bea}{\begin{eqnarray}}
\newcommand{\eea}{\end{eqnarray}}
\newcommand{\beq}{\begin{equation}}
\newcommand{\eeq}{\end{equation}}

  \newcommand{\half}{{\small 1\over 2}}
  \renewcommand{\>}{\rangle}

\renewcommand\ell{l}

\newcommand{\ndim}[1]{#1-d}

\usepackage{placeins}

\newcommand\Pm{P_-}
\newcommand\Pp{P_+} 
\newcommand\Pmg{\mathbb{P}}

\newcommand\BNO{{\tt BNO}}

\title{Multigrid for Chiral Lattice Fermions: Domain Wall}

\author[*]{Richard C. Brower}
\author[$\dagger$]{M. A. Clark}
\author[*]{Dean Howarth}
\author[$\dagger$]{ Evan S. Weinberg}
\affil[*]{Boston University, Boston, MA 02215, USA}
\affil[$\dagger$]{NVIDIA Corporation, Santa Clara, CA 95050, USA}

\date{\today}
\begin{document}

\maketitle

\begin{abstract}

{\em Critical slowing down} for the Krylov Dirac solver presents a
  major obstacle to further advances in lattice field theory as it
  approaches the continuum solution. We propose a  
  new multi-grid approach for chiral fermions, applicable to both the \ndim{5} domain wall or \ndim{4}  Overlap operator. 
  The central idea is to directly coarsen the \ndim{4} Wilson kernel, giving
  an effective domain wall or overlap operator on each level.
We provide here an explicit construction for the Shamir domain wall formulation with numerical tests for the   
 \ndim{2} Schwinger prototype, demonstrating near ideal  multi-grid scaling. The framework is designed
  for a natural extension to \ndim{4} lattice QCD chiral fermions,  such as  
  the  M\"obius,   Zolotarev  or Borici domain wall discretizations
  or directly to a  rational expansion  of the \ndim{4}  Overlap operator.
  For the Shamir operator, the effective overlap operator is
  isolated  by the use of a Pauli-Villars preconditioner
 in  the  spirit of the K\"ahler-Dirac spectral map used in a recent 
  staggered MG algorithm~\cite{Brower:2018ymy}. 
  \end{abstract}

\setlength{\parskip}{0in}
\thispagestyle{empty}
\setcounter{page}{-1}
\pagebreak
\tableofcontents
\setcounter{page}{1}
\setlength{\parskip}{.2in}

\newpage
\section{\label{sec:intro}Introduction}

Increasingly powerful computers and better theoretical insights
continue to improve the predictive power of lattice quantum field
theories, most spectacularly for lattice quantum chromodynamics
(LQCD)~\cite{Wilson:1974sk}.  However, with larger lattice volumes and
finer lattice spacing exposing multiple scales, the  lattice Dirac solver
becomes increasingly ill-conditioned threatening further progress.  The cause is well known: as the
fermion mass approaches zero the Dirac operator becomes singular due
to the exact {\em chiral} symmetry of the Dirac equation at zero
mass, causing {\em critical slowing down}~\cite{Blum:2013mhx}.  The
algorithmic solution to this problem for lattice QCD was recognized 30
years ago~\cite{Brower:1990at}: the fine grid representation for the linear solver should
be coupled to multiple scales on coarser grids in the spirit of Wilson's real space
renormalization group and implemented as a recursive multi-grid (MG)
preconditioner~\cite{Wilson:1974mb}. Early investigations in the 1990's
introduced a gauge-invariant productive MG algorithm~\cite{Brower:1991xv,Hulsebos:1990er} with encouraging results for
the Dirac operator in the presence of weak (or smooth) background
gauge fields near the continuum.  However, in practice lattice sizes
at that time were too small and the gauge fields were too rough to
achieve useful improvements.

It was not until the development of adaptive geometric MG
methods~\cite{Brannick:2007ue, Babich:2010qb} that a fully recursive
MG algorithm, capable of projecting strong background chromodynamics fields
onto coarser scales, was found for the Wilson Dirac discretization. However there are two other discretizations, referred to as staggered~\cite{PhysRevD.11.395} and domain wall~\cite{Kaplan:1992bt}
fermions, that are  used extensively in high energy
applications that  more faithfully represent {\em chiral} symmetry on
the lattice. The extension of adaptive geometric MG to these discretizations has
proven to be more difficult, perhaps related to the improved lattice
chiral symmetry. There has been progress on a two-level MG algorithm for domain wall fermions~\cite{Cohen:2012sh,Boyle:2014rwa,Yamaguchi:2016kop} and a non-Galerkin algorithm for the closely related overlap operator~\cite{Neuberger:1997bg,Brannick:2014vda}. Recently an adaptive geometric multigrid algorithm for staggered fermions was discovered based on a novel pre-conditioner inspired by the K\"ahler-Dirac
spin structure~\cite{Becher1982,PhysRevD.38.1206}. 

Here we propose a new approach to the domain wall
discretization which leverages, at least on a heuristic level, features developed from both the Wilson and staggered MG methods. We hope that a comparison of methods will lead to new optimizations across the full set of discretizations. The design strategy of our domain wall MG algorithm 
consists of trying to separate the \ndim{4} physical subspace of low modes found in the effective \ndim{4} overlap operator from the larger \ndim{5} domain wall vector space. This procedure
is conveniently enumerated in 3 steps:
\begin{itemize}[noitemsep,topsep=0pt]
\item [ i.]   {\bf Approximate Pauli-Villars preconditioning} of the domain wall operator~\cite{Brower:2012vk}.
\item[  ii.]  {\bf Wilson kernel MG projection} on the
domain wall and Pauli-Villars factors~\cite{Babich:2010qb}.
\item[iii.] {\bf Truncated projection/prolongation} restricted  to the
domain wall boundary.
\end{itemize}

The salient features of each step are:
{\bf i.)} The exact Pauli-Villars inverse $D^{-1}_{PV}$, which is a perfect  map
from the DW spectrum onto overlap, is well approximated  by the
application of  the  Pauli-Villars adjoint, $D^\dag_{PV}$. {\bf ii.)} A Galerkin MG
coarsening is applied to the \ndim{4} Wilson kernel on each extra-dimensional slice separately for both the domain
wall and Pauli-Villars factors. The null space projection is formulated entirely from the \ndim{4} Wilson kernel and does not scale with the size of the extra dimension. {\bf iii.)} Finally, within the multigrid cycle, the residual coarsening and error interpolation is restricted to the domain wall, which in turn allows the extent of the extra dimension of the coarse-level operator to be reduced. 

We again follow the successful development strategy for the
Wilson~\cite{Brannick:2007ue} and staggered~\cite{Brower:2018ymy} MG algorithms by using the two-flavor lattice Schwinger model~\cite{PhysRev.128.2425,Smilga:1996pi} as a prototype for exploration and testing. The reader is referred to Fig.~\ref{fig:iterationscompare}  and the accompanying  Table~\ref{sec:DWformalism} for a concise summary of the performance of our domain wall algorithm for the \ndim{2}  Schwinger model.
In Sec.~\ref{sec:DWformalism}, the underlying motivation and formalism is given. For simplicity, the discussion is restricted to the Shamir
domain wall operator~\cite{Shamir:2000cf}. This is followed in Sec.~\ref{sec:Results} by the details of the implementation and benchmarks for our prototype \ndim{2} Schwinger model.
Care is taken to present the formalism in a dimension-agnostic form to accommodate  extensions from \ndim{2} to \ndim{4} gauge theories. In Sec.~\ref{sec:Conclusion} we conclude by noting that our core developments not only apply to the \ndim{4} Shamir formulation presented here but also to the M\"obius~\cite{Brower:2004xi,Brower:2005qw}, Borici~\cite{Borici:1999da,Borici:1999zw}, and Zolotarev~\cite{Chiu:2002ir,Chiu:2002kj} formulations, as well as directly to the overlap operator approximation to the sign function~\cite{Neuberger:1997bg,Edwards:2000rk,Edwards:2000qv,Edwards:2005an}.

\section{\label{sec:DWformalism}Domain Wall Formalism}

All lattice discretizations of the Dirac operator seek to rapidly approach the continuum Dirac operator,
\be
D \psi(x) =  \gamma_ \mu (\partial_\mu - i A_\mu(x))\psi(x)  + m \psi(x) \; ,
\label{eq:continuum}
\ee
as the lattice spacing vanishes.  The continuum operator is a first derivative, an anti-Hermitian operator, thus the spectrum is imaginary indefinite except for a small real shift for $m > 0$. It obeys an exact chiral symmetry at zero mass ($m = 0$). The Wilson discretization, 
\be
D_W\left(U,m\right)_{x,y} =  - \frac{1-\gamma_\mu}{2}U_\mu(\vec{x})\delta_{x+\mu,y} -  \frac{1+\gamma_\mu}{2}U^\dagger_\mu\left(\vec{x}-\hat\mu\right)\delta_{x-\mu,y}+\left(d+m\right)\delta_{x,y} \; ,
\ee
introduces an anti-Hermitian ``na\"ive'' first difference and adds a Hermitian second-difference (or so-called Wilson term) to lift the
doublers to the cut-off scale $\pi/a$ at the expense of explicitly violating lattice chiral symmetry at $\mathbb{O}(a)$ in lattice spacing. The Wilson lattice operator then requires fine tuning of the bare quark in order to restore chiral symmetry in the continuum limit.

The chiral overlap~\cite{Neuberger:1997bg, Neuberger:1997fp} and domain wall~\cite{Shamir:2000cf}
fermions, beyond their remarkable physical properties, have the feature that the Wilson kernel can be re-purposed. In the domain wall (DW) approach, the Wilson kernel is present in an extra dimension separating \ndim{4} domain walls by a lattice of length $L_s$. Suppressing the four-dimensional indices, the DW operator is given by
\begin{align}
D_{DW}\left(m\right)_{s's} &= \left[\begin{array}{ccccc}
D_W(M_5) + 1 & P_{-} & 0 & \cdots & -mP_{+} \\
P_{+} & D_W(M_5)+1 & P_{-} & \cdots & 0 \\
0 & P_{+} & D_W(M_5)+1 & \cdots & \vdots \\
\vdots & \vdots & \vdots & \ddots & P_{-} \\
-mP_{-} & 0 & \cdots & P_{+} & D_W(M_5)+1
\end{array}\right]
\label{eq:DWoperator}
\end{align}
where $P_{\pm} = \frac{1}{2}\left(1 \pm \gamma_5\right)$. The  indices $s,s' = 1,\cdots L_s$ label \ndim{4} blocks in the extra fifth dimension (or d+1 dimension). The bulk mass, $M_5 < -1$, is tachyonic. The physical bare mass parameter is encoded by the boundary parameter $m$.

In the limit of $L_s \rightarrow \infty$ an exact lattice chiral symmetry appears up to an explicit fermion mass gap given by
\begin{align}
m_q = \frac{m}{1 - m} \simeq m \quad \mbox{as} \quad m \rightarrow 0 \; .
\end{align}
The result is that propagators between the domain walls are described by the effective \ndim{4} overlap operator proposed by
Neuberger~\cite{Neuberger:1997bg,Kikukawa:2000ac} with the deformed chiral algebra of the Ginsparg-Wilson identity~\cite{Ginsparg:1981bj},
\be
\gamma_5 D^{-1}_{ov} +  D^{-1}_{ov} \gamma_5  = \mathcal{O}( a ) \; ,
\ee
at zero quark mass. The explicit spectral map from  the domain wall
to the overlap operator  will be presented in  Sec.~\ref{sec:specmap}, following closely the notation in~\cite{Brower:2012vk} which we will refer to as \BNO. This spectral map between domain wall and overlap operators plays a central role in our DW MG algorithm.

\begin{figure}[t] \centering
    \includegraphics[width = 0.9\textwidth]{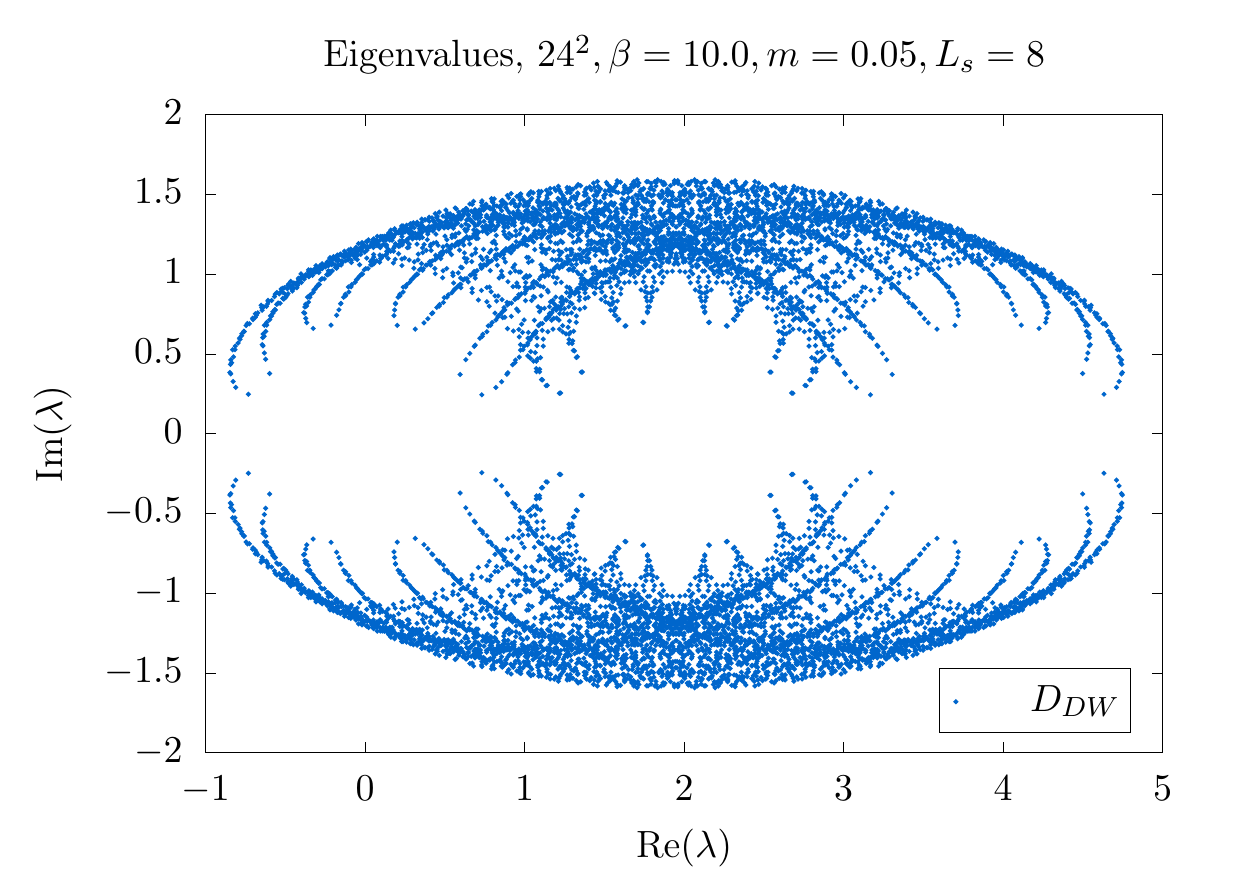}
    \caption{The spectrum of the domain wall operator lacks the continuum chiral low modes; instead there are $2^{d/2} L_s$ of them ($2^{d/2}$ coming from the spin degrees of freedom) in a circle around zero in the complex plane.}
    \label{fig:dw_spectra}
\end{figure}

\subsection{Standard Approaches and Shortcomings} 

The domain wall operator encodes chiral symmetry in a subtle and indirect
fashion. The full spectrum of the domain wall operator in Eq.~\ref{eq:DWoperator}, illustrated in Fig.~\ref{fig:dw_spectra} for two dimensions, {\bf does not have the expected small eigenvalues of the continuum as you approach the chiral limit}, but instead has $\mathcal{O}(L_s)$ small eigenvalues. This can easily be seen in the free limit ($U = 1$)  which at zero momentum has the spectrum,
\be
\lambda_n = 
m^{1/L_s} e^{\; \textstyle  i(2n+1) \pi/L_s}\quad \mbox{for} \quad  n = 0, 1, ..., L_s 
-1 \; .
\ee
In the exact chiral limit,  first  taking $L_S \to \infty$  followed by
$m \to 0$, this operator has no zero modes. Instead they form a unit circle  around the  origin in the complex plane. This feature persists when gauge fields are turned on
as illustrated in Fig.~\ref{fig:dw_spectra} for two dimensions with Abelian gauge fields.

The DW operator features further issues. Unlike with staggered or Wilson fermions, the DW operator is dramatically non-normal ($[D_{DW}(m),D^\dagger_{DW}(m)] \ne 0$), even in the exact free field. The spectrum does not satisfy the half-plane eigenvalue condition with positive real values ($Re[\lambda] > 0$). These two properties defeat reliable convergence properties of traditional Krylov solvers. For example, consider a normal  indefinite matrix, whose spectrum fits in a circle of radius $r$ centered at a complex point $c \in \mathbb{C}$. For GMRES methods, one can show the relative residual on iteration $n$ of a GMRES method is bounded by $\left|r/c\right|^n$~\cite{doi:10.1002/gamm.201490008}.

A standard method to solve the domain wall linear system is to replace
\be
D_{DW}(m)~\Psi = b
\ee
with the normal system,
\be
D^\dag_{DW}(m) \; D_{DW}(m)~\Psi = D^\dag_{DW}(m)~b \;.
\ee
This system has multiple benefits but also complications. 

The normal operator encodes a single low mode with, in the free-field limit, eigenvalues $\mathcal{O}(m^2)$. The low modes are ``bound'' to the domain wall, as can be visually inspected by looking at the profile of the eigenvectors (the singular vectors of the domain wall operator) in
the bulk dimension. Both of these properties form a stark contrast with $D_{DW}$;  the normal operator transforms  the physics of the domain wall operator into as a single chiral fermion below the cut-off.  

From a numerical standpoint, the normal operator is Hermitian positive definite (HPD) and can be solved efficiently by traditional Krylov methods, e.g., Conjugate Gradient. Further, it is amenable to deflation with eigenpairs generated via an efficient Lanczos process. Also as a Hermitian positive definite matrix, the solver for this normal operator can be implemented as a traditional MG algorithm~\cite{Cohen:2012sh,Boyle:2014rwa}. However, a numerical implementation of the coarsened normal operator, a distance-two stencil, requires a non-trivial increase in computation and communication relative to a distance-one stencil~\cite{benson1973iterative,Chow:2001:PIP:1080623.1080641,Sterck06reducingcomplexity,doi:10.1137/140952570}. This owes in large part to a far more complicated gather pattern due to around-the-corner terms, which the original fine level original normal operator avoids by being the product of two distance-one operators.

One solution to this issue is to recognize that this normal operator is the square of a distance-one ``$\gamma_5$'' Hermitian  operator,
\be
D^\dag_{DW}(m) \; D_{DW}(m) = (\Gamma_5  D_{DW}(m))^2 \; ,
\ee
where  $\Gamma_5 =\gamma_5 R $ is a product of $\gamma_5$ and the reflection in the extra dimension by $R_{ss'} =  \delta_{(s+ s'-1) \% L_s, 0}$. This operator  has an indefinite real spectrum similar to the imaginary  
spectra in the continuum and staggered operator on the lattice. Appealingly, the operator also has  a single chiral mode with eigenvalues of $\mathcal{O}(m)$. However, this operator itself has its own frustrations: while it can be coarsened, it develops spurious small eigenvalues similar to with a na\"ive approach to staggered fermions, which was shown in~\cite{Brower:2018ymy} to harm a fully recursive algorithm.

The $\Gamma_5 D_{DW}$ operator leads itself to a clean interpretation of spurious low modes via a free-field analysis. The low modes of $\Gamma_5 D_{DW}$ do not include support on the bulk; as such, a ``coarsening'' inspired by the low modes eliminate it. In this case, the coarsened operator can be shown to be $\gamma_5 D_{\mbox{\small{na\"ive}}}$ with well understood extra low modes. This is actually a foretelling of a salient feature of our algorithm: the bulk dimension cannot be trivially eliminated. As a last concern, this approach does not generalize beyond Shamir domain wall fermions: the fully general M\"obius formulation has a Wilson kernel inverse as part of its definition of $\Gamma_5$. Our new approach
seeks to avoid these difficulties base on the spectral map form the domain wall and to the overlap representations in \BNO, combined  with methods borrowed 
from  prior multigrid algorithms for Wilson~\cite{Brannick:2007ue} and staggered~\cite{Brower:2018ymy} discretizations.

\subsection{Spectral Map from Domain Wall to Overlap}
\label{sec:specmap}

One may view the Pauli-Villars operator, $D_{PV}\equiv D_{DW}(1)$
as a left preconditioner of the domain wall operator in the linear system
\be
 D^{-1}_{PV}  D_{DW}(m) \Psi =  D^{-1}_{PV} b  \; .
\ee
We will show that the Pauli-Villars operator is an ideal, albeit
expensive, preconditioner and that even a simple approximation to the inverse dramatically accelerates convergence.  This is
accomplished  via the generalized eigenmode problem, $ D_{DW}(m)\Psi_\lambda  =  \lambda D_{PV} \Psi_\lambda$, which separates
the low chiral generalized eigenvectors  ($\lambda \simeq 0$) bound to the walls at $s = 1$ and $s = L_s$,
\be
\psi_x =  \half(1 - \gamma_5)   \Psi_{x,1}  + \half(1 + \gamma_5)  \Psi_{x,L_s},
\ee
from the high bulk modes at the cut-off: $\lambda = \mathcal{O}(\pi/a)$.

To see this  explicitly, it is convenient as in {\tt BNO} to first move
both walls to $s = 1$ by introducing a cyclic permutation
of the  negative chiral modes at $s = L_s$ to $s=1$
by 
\begin{equation}
 {\cal P}_{s's} =  \left[ 
{\begin{array}{cccc}
\Pm & \Pp & \cdots & 0 \\
0 & \Pm & \Pp \cdots &0 \\
\vdots & \vdots & \ddots & \vdots \\
0 & 0 & \cdots & \Pp\\
\Pp & 0 & \cdots & \Pm
\end{array}}
 \right]_{s's} \; ,
\end{equation}
with $P_{\pm} =\frac{1}{2} (1 \pm \gamma_5)$. This defines a unitary transformation of our domain wall operators,
\be
 D_{DW}(m)   \rightarrow  {\cal P^\dag } D_{DW}(m) {\cal P}.
\ee
Following the derivation via LDU transformation in \BNO,  the preconditioned matrix in this {\em permuted chiral basis},
\begin{equation}
K_{DW}(m) = ( D_{PV} {\cal P})^{-1} D^{DW}(m) {\cal P}
 = \left[ \begin{array}{rrrrrrr}
D_{ov}(m) & 0 & 0 & \cdots & \cdots &  0\\
-(1-m)  \Delta_{2}  & 1 &0 &0&\cdots & 0\\ 
-(1-m)\Delta_{3} & 0 &1 & 0 & \cdots &0\\
-(1-m)\Delta_{4} & 0 & 0 & 1  & \cdots & 0 \\
\vdots & \vdots &  \ddots & \ddots &\ddots  &\vdots \\
-(1-m)\Delta_{L_s} & 0 & \cdots &\cdots&0& 1
\end{array}\right]
\label{eq:DWFmatrix} \; .
\end{equation}
is mapped into a block diagonal form~\cite{Brower:2012vk} in the extra dimension. This remarkable identity is the central observation for our preconditioned
multigrid algorithm. The effective overlap operator block, $K^{DW}_{1,1}(m)=D_{ov}(m)$, has 
all the non-trivial low  eigenvalues. The  additional extra heavy modes are mapped 
{\bf exactly} to unit eigenvalues (or in physical units at the $1/a$), irrespective of $L_s$, lattice spacing (here scaled to $a = 1$) and gauge interactions. 

Parenthetically,  we should acknowledge  that this general mechanism to isolate the
low domain wall modes in an effective overlap block is well known and it
is the fundamental insight to chiral lattice fermions~\cite{Neuberger:1997fp}. In particular the Monte Carlo sampling of the path integral  must divide by
the Pauli-Villars determent  to give a finite determinant ratio, $\det[K] = \det[ D_{DW}(m)]/\det[D_{DW}(1)]$ in the continuum. As explained by Kaplan and Schmaltz in~\cite{Kaplan:1999jn} in
an elegant exposition  based  on a {\it kinematic  super symmetry} cancellation between the bulk fermion and the bosonic Pauli-Villars pseudofermions,  broken only by domain wall boundary, to
give rise to boundary chiral modes via the Callan-Harvey descent relations~\cite{Callan:1984sa}. 

\begin{figure}[t]
    \centering
   \includegraphics[width=0.9\linewidth]{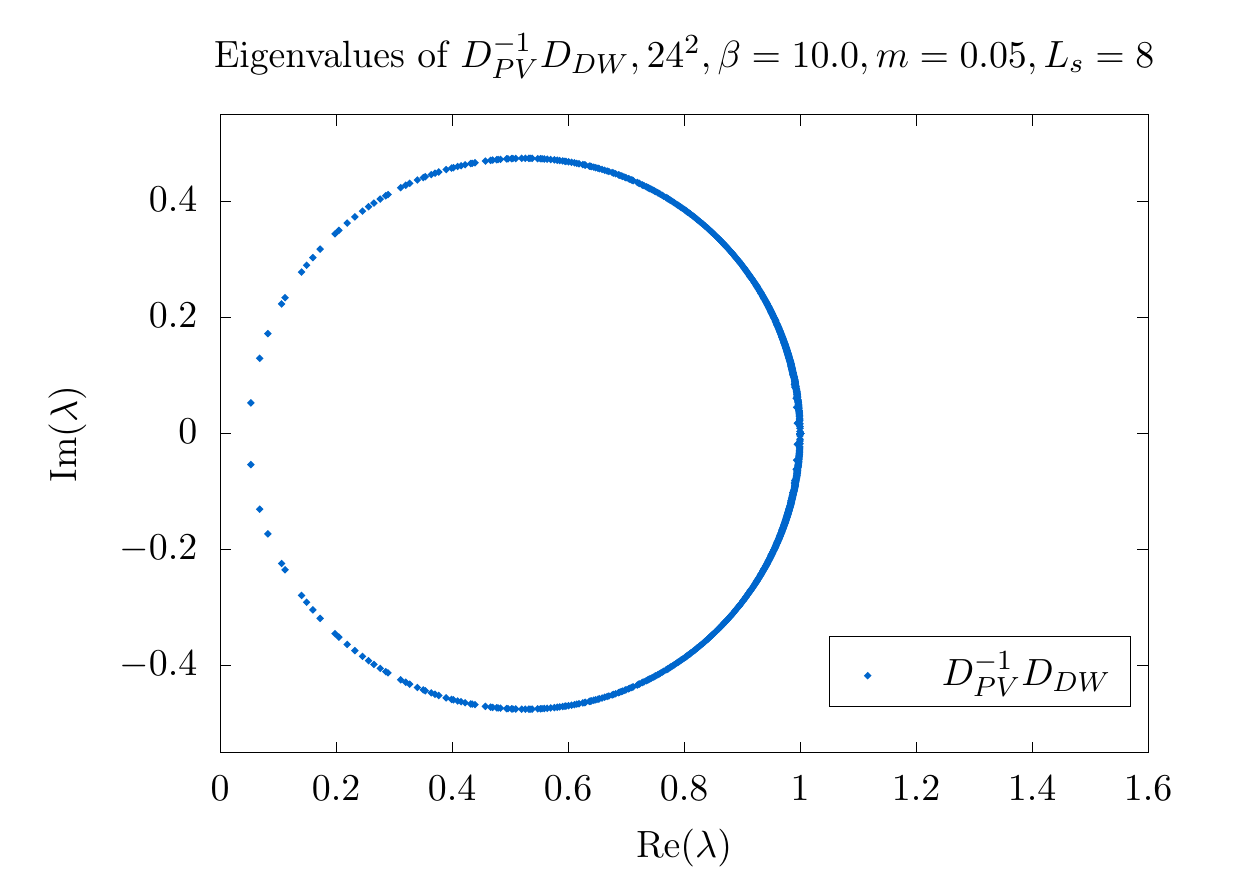}  
    \caption{\label{fig:overlap_spectrum}A representative spectrum of the preconditioned domain wall operator, $D_{PV}^{-1} D^{DW}(m)$. Deviations from an exact circle exist but are qualitatively negligible. Note the similarity to the K\"ahler-Dirac preconditioned staggered operator in Fig.~\ref{fig:freespecStag}.} 
\end{figure}

The block structure of $K_{DW}(m)$ lends itself to a structured block-inverse given by
\begin{equation}
K^{-1}_{DW}(m)  
 = \left[ \begin{array}{rrrrrrr}
D^{-1}_{ov}(m) & 0 & 0 & \cdots & \cdots &  0\\
(1-m)  \Delta_{2}  & 1 &0 &0&\cdots & 0\\ 
(1-m)\Delta_{3} & 0 &1 & 0 & \cdots &0\\
(1-m)\Delta_{4} & 0 & 0 & 1  & \cdots & 0 \\
\vdots & \vdots &  \ddots & \ddots &\ddots  &\vdots \\
(1-m)\Delta_{L_s} & 0 & \cdots &\cdots&0& 1
\end{array}\right]
\label{eq:Kinvmatrix} \; .
\end{equation}
The identification of the overlap propagator $G_{11} \equiv D^{-1}_{ov}(m)$ agrees with the practice of computing DW propagators by solving the linear system, $D_{DW}(m) G   = D_{PV} b $, from an arbitrary source on the wall $b \sim \delta_{1,s}$.
The heavy modes $(0, 0, \cdots , 1,\cdots,0)$ are static. On the other hand, the non-zero elements in the first column show that the chiral modes bleed exponentially into the interior by factors, $\Delta_{s+1} =  T \Delta_{s}  = T^s/(1 + T^L)$ in terms of the transfer matrix: $T = (1 -H)/(1 + H)$. At finite  $L_s$ the overlap operator is 
\be
D_{ov}(m) =\frac{1 + m }{2}+ \frac{1 -m }{2} \gamma_5 \epsilon_L[H] = m + (1-m) D_{ov}(0),
\label{eq:Overlap}
\ee
with 
\be
\epsilon_L[H] = \frac{(1 - H)^L - (1 + H)^L}{(1 - H)^L +  (1 + H)^L} \quad \mbox{and} \quad  H = \gamma_5 D_W(M_5)/( 2 + D_W(M_5))
\ee
in the Shamir implementation. In the limit $L_s \rightarrow \infty$ this becomes the exact sign function:  $ \epsilon_L[H] \rightarrow \widehat \gamma_5 = \mbox{\tt sign}[H]$. The  same spectral transformation into this sparse  structure in Eq.~\ref{eq:Kinvmatrix} applies to  other implementations of domain wall fermions (M\"obius, Borici,  Zolotarev, etc). The modifications include a variation of the Hermitian kernel $H$ and the functional $\epsilon_L[H]$ that converge, $\epsilon_L[H]\rightarrow \mbox{\tt sign}[H] $ to the sign function
as $L_s \rightarrow \infty$. 

We can trace the sparse  block structure to the mass dependence on the dyadic
structure at the boundaries relating the  Pauli-Villars operator to the domain wall operator,
\be
D_{PV} =  D_{DW}(m) + (1-m) 
\begin{bmatrix}
P_+ \\
0  \\
0  \\
  \vdots \\ 
 P_- 
\end{bmatrix}
\otimes
\begin{bmatrix}
P_- & 0  & 0 & \cdots & P_+ \\
\end{bmatrix}
\ee
or $D_{PV} \equiv D_{DW}(m) + (1-m) U V^\dag$. After applying  the  Sherman-Morrison-Woodbury formula~\cite{golub13}, 
\be
D^{-1}_{PV} D_{DW}(m) = 1   - D^{-1}_{DW}(m) U \frac{(1-m)}{ I + (1-m)V^T  D^{-1}_{DW}(m) U } V^T  \; ,
\ee
and again considering the chiral basis, $V^T \rightarrow V^T {\cal P}
=
\begin{bmatrix}
1 & 0  & 0 &   \cdots & 0 \\ 
\end{bmatrix} $, 
we see a clear projection onto the first column in $K^{DW}_{s,s'}(m)$, reproducing the sparse structure in Eq.~\ref{eq:DWFmatrix}.

\paragraph{Free-Field Limit:} The  analysis of the free-field ($U =1$)  limit
for the domain wall and Pauli-Villars operators  
gives  valuable insight and guidance to our MG construction, particularly
when examining the low spectra well below the UV cut-off scale, $\pi/a$. Qualitative
and even quantitative features survive the introduction of the gauge fields
generated by lattice Monte Carlo methods.

\begin{figure}[th]
    \centering
 \includegraphics[width=0.9\linewidth]{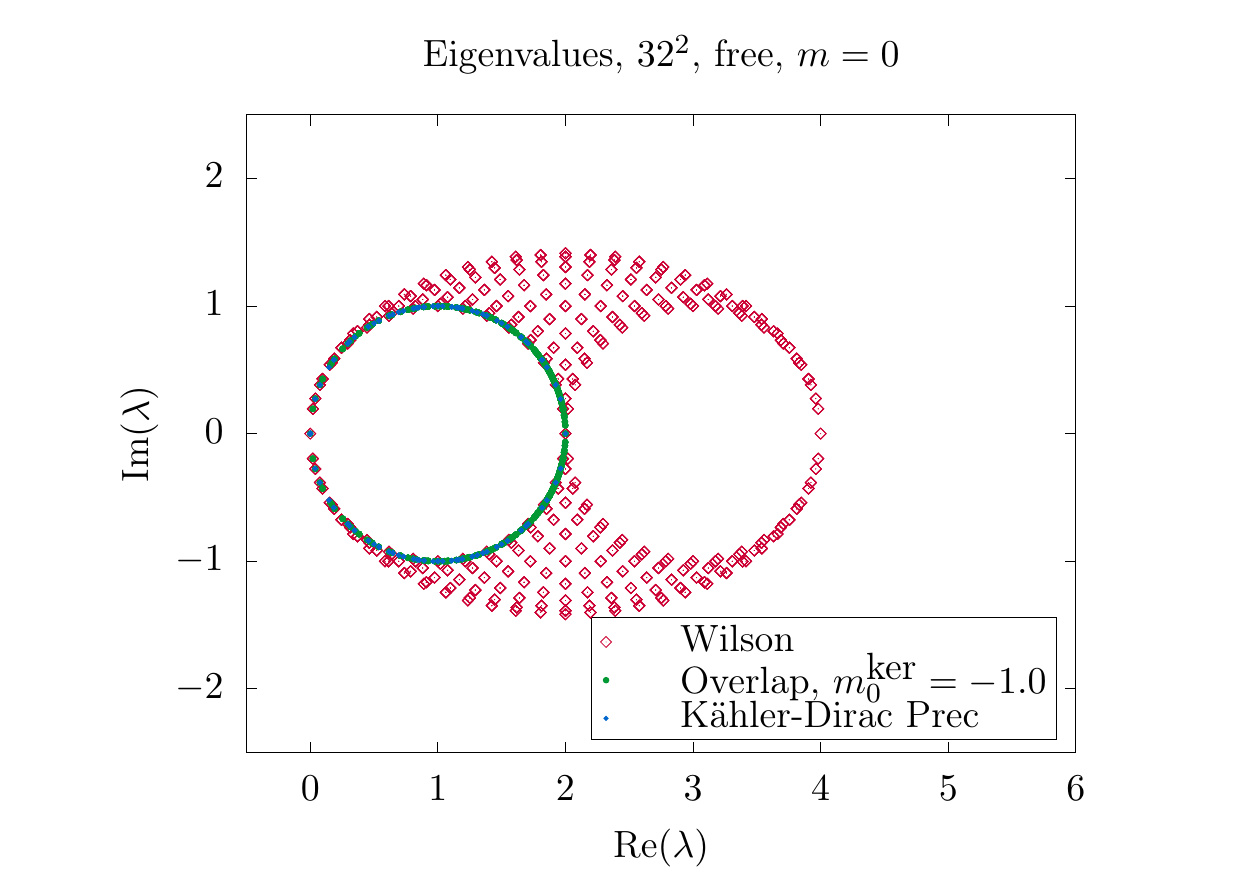}
   \caption{\label{fig:freespecWilson}
   Comparison of the 2-d free effective overlap and   K\"ahler-Dirac preconditioned operator with almost identical
   circles centered at $\lambda =  1$ overlaid with 2-d free Wilson operator with  two doublers at 
   $\lambda = 2$ and one at $\lambda =4$.}
\end{figure}

We  begin by transforming the free  Wilson kernel in Eq.~\ref{eq:DWoperator} to momentum space, 
\be
\widetilde D_W( p_\mu) = a m + \gamma_\mu \sin( a p_\mu) + 2  \sin^2(a p_\mu/2)
= a m + \sum_\mu (1 - e^{\textstyle -i \gamma_\mu a p_\mu}) \; .
\ee
where we introduce the expression on the right to emphasize the well known feature of circular arcs in the complex spectrum of the Wilson operator. This gaping separates the doubler modes from the continuum modes as evident in Fig.~\ref{fig:freespecWilson}
even with  non-zero gauge fields  turned on. The lattice spacing ${\bf a}$ has been introduced to identify
 physical low modes ($|p| \ll  \pi/a$) relative to UV cut-off: $\mathcal{O}(\pi/a)$. The low
spectrum,  $\lambda_\pm \simeq  m +  \pm  i \sqrt{p^2} + a p^2/2$, is the continuum  Dirac spectrum  plus the $\mathcal{O}(a p^2)$ Wilson term.

This Fourier analysis had a straightforward generalization to the Pauli-Villars operator, $D_{PV}$, as the boundary conditions are antiperiodic in the fifth dimension,
\be
\widetilde D_{PV}(p_\mu,p_5) =  -   e^{\textstyle -i \gamma_5 a p_5} + \widetilde D_W( p_\mu) +  (1 + M_5) \; .
\ee
The boundary conditions restrict the bulk momentum $p_5$ to half integer modes: $p_5 = \pi( 2 n +1)/L_s$.
After setting the mass to its free-field tachyonic value $M_5 = -1$,  the low momentum expansion 
 again has the familiar Wilson form  ($ i \gamma_\mu p_\mu + (a/2) p^2 $) with
 first ``eye'' displaced to a circle centered at $\lambda = 1$. Not surprisingly
 the free domain wall operator,  $D_{DW}(m)$, which differs by a fermion-mass dependence on the boundary (reducing to Dirichlet boundary conditions for $m = 0$), has a qualitatively similar spectra as seen in Fig.~\ref{fig:dw_spectra} even with non-zero gauge fields. Both have {\bf no small} eigenvalues below the cut-off for free fields. (With interactions
small eigenvalues occur when the   topological charge changes.) 

Turning to the normal equation, we see a dramatic difference.
While the domain wall normal operator has chiral modes at $m=0$, as we show
in Appendix~\ref{app:lowepequiv}, the  normal equation for the Pauli-Villars operator is positive definite
with a large gap from zero with {\bf singular values} of order the cut-off:
\be 
D_{PV} D_{PV}^\dagger = 1+ \mathcal{O}(p^4) \label{eq:loworderpv}
\ee
The  first correction is $\mathcal{O}(a^2p^4)$ in physical units. 
Indeed more generally for $M_5 = -1$ one can prove for any finite $L_s$ in the free limit 
that $D_{PV}^\dagger D_{PV}$ is bounded from below by 1, i.e., at the lattice cut-off scale: $1/a^2$. This feature suggest the usefulness of our approximate preconditioning,
\be
D_{PV}^{-1}D_{DW} = D^\dag_{PV} (D_{PV} D_{PV}^\dagger)^{-1}  D_{DW}  \simeq D_{PV}^{\dag}D_{DW} \; 
\ee
which avoids the expensive need to invert the Pauli-Villars operator.
Even after gauge fields
are included, we note  in Fig.~\ref{fig:overlap_and_approx}  this approximation  conforms well at small
eigenvalues including the $\mathcal{O}(a p^2)$ for the parabolic curvature in the complex plane. 

To explore this further, we summarize results from Appendix~\ref{app:lowepequiv}, comparing the low momentum spectra 
for the exact overlap map, $D_{PV}^{-1} D_{DW}$, with the approximate map, $D_{PV}^{\dag} D_{DW}$.  At $m = 0$, the low spectrum for $D_{PV}^{-1} D_{DW}$ is given by $\lambda^0_\pm =   \pm i\sqrt{p^2} + a p^2  + \mathcal{O}(a^2)$. By  using the shift identity,
$\lambda^0 \rightarrow \lambda = a m + (1 - a m)\lambda^0$ as implicit in Eq.~\ref{eq:Overlap}, the low spectrum for non-zero mass is
\begin{align}
\lambda_\pm =   a m \pm  i (1-a m)\sqrt{p^2} + a (1-a m) p^2  + \mathcal{O}(a^2).
\label{eq:ExactLowEV}
\end{align}
which after a rescaling, $\lambda_\pm \rightarrow \lambda_\pm/(1-a m) = a m_q \pm  i \sqrt{p^2} + a  p^2$,
is a Wilson-like dispersion relation.  The general expansion
in powers of momentum $p^2$ has a rather remarkable independence on $L_s$.  As noted in Appendix~\ref{app:lowepequiv}, a direct evaluation of the free overlap kernel in Eq.~\ref{eq:Overlap}  for  any finite $L_s \ge 2$ results in 
a series in $p^2$ which, up to $\mathcal{O}((a^2 p^2)^k)$m is independent of  $L_s \ge k$. This invariance of the low momentum expansion with respect to size of the extra dimension may explain the efficacy 
of reducing the size of extra dimension $L_s$
on the coarse level iterations as documented in Sec.~\ref{sec:redcoarse}.

Finally we  compare the low mass spectra to our approximate preconditioned operator, $D^\dag_{PV} D_{DW}(m) $, 
\be
\lambda_\pm   =  a m \pm  i (1-a m) \sqrt{p^2} + a p^2 + \mathcal{O}(a^2) \; .
\label{eq:ApproxLowEV}
\ee
relative to the exact preconditioner in Eq.~\ref{eq:ExactLowEV}. The only difference to quadratic order occurs at dimension 6 with a contribution
$\mathcal{O}(a^2 m p^2)$, helping to explain why a Wilson-like spectra in the overlap sector is
preserved in Fig.~\ref{eq:coarsenDW}, including the parabolic curvature to $\mathcal{O}(a p^2)$.

\begin{figure}[th]
    \centering
    \includegraphics[width=0.9\linewidth]{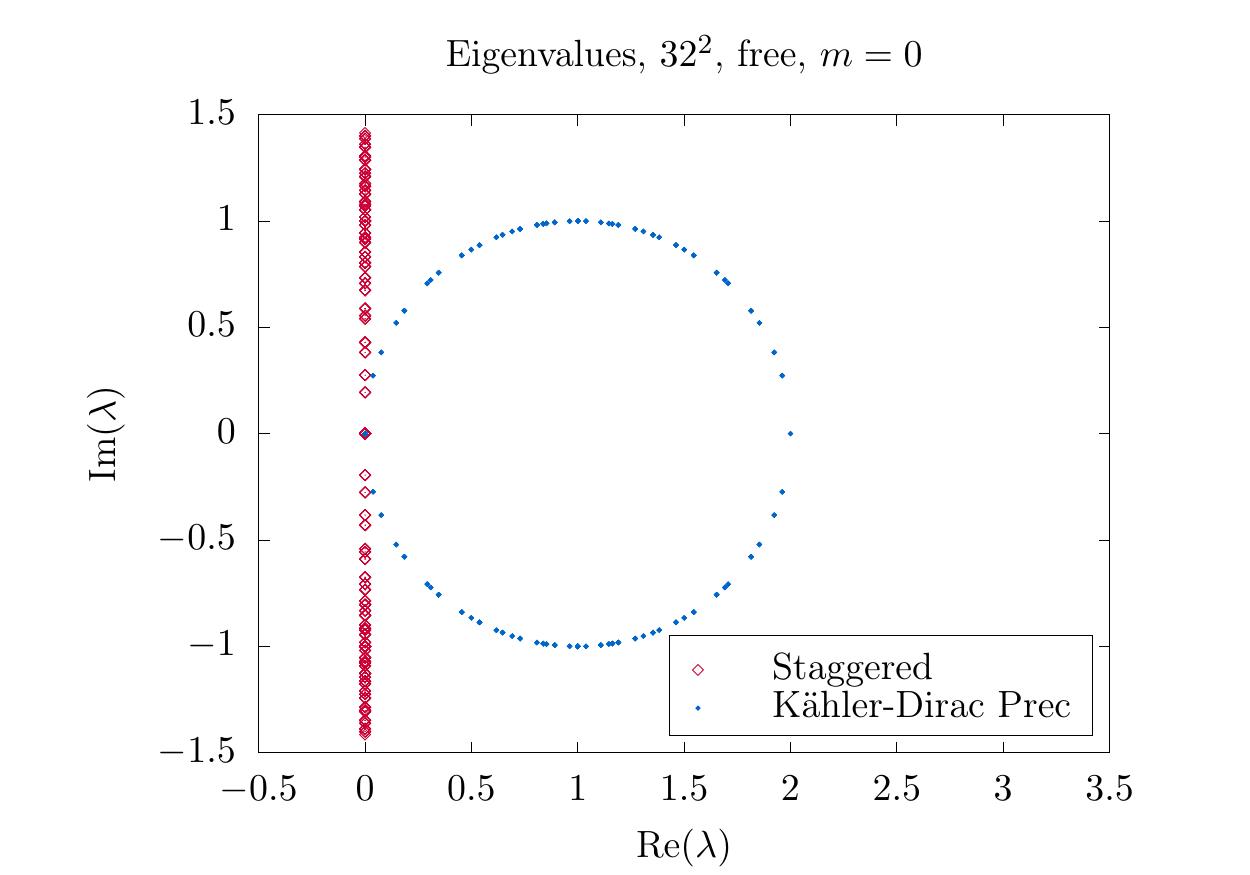}
   \caption{\label{fig:freespecStag}The spectrum of the two-dimensional free, massless staggered operator before (vertical line) and after K\"ahler-Dirac preconditioning.}
\end{figure}

\paragraph{Similarity with the K\"ahler-Dirac Preconditioned Staggered Operator:}
It is interesting to compare the Wilson and the overlap
spectra with the preconditioned spectrum for the staggered MG algorithm in Ref.~\cite{Brower:2018ymy}. The staggered lattice operator has the unique property, shared by the 
continuum, of being an exactly anti-Hermitian operator plus a constant mass shift as illustrated for $m =0$ by the vertical (red) spectra in Fig.~\ref{fig:freespecStag}. Both the staggered and continuum operators are normal operators.

At first these similarities between the staggered operator and the continuum may seem to be ideal for multigrid, but this turned out to be major obstacle to extending the Galerkin projection method used successfully for the Wilson MG algorithm~\cite{Brannick:2014vda} to the staggered operator. The solution found in Ref.~\cite{Brower:2018ymy} was to first precondition by {\em dividing} the anti-Hermitian staggered operator by the spin-taste  K\"ahler-Dirac block, deforming
the spectrum into one resembling the overlap spectrum or the first
``eye'' of Wilson spectra  in  Fig.~\ref{fig:freespecStag}. 

More specifically, this required writing the staggered operator as an ``even/odd'' $2^d$ block operator (i.e., $2^2$ squares in 2d, $2^4$ hypercubes in 4d) decomposed as a sum of block-local terms ``$B$'' and block-hopping terms ``$C$'' as described in Eq. 2.11 of~\cite{Brower:2018ymy}. Each of these terms are separately anti-Hermitian  operators with an indefinite spectra. In this formalism, the preconditioned operator is simply the block Jacobi preconditioned form, $B^{-1}~D_{stag}$. This maps the spectrum onto an exactly unitary circle in the free case, as shown in Fig.~\ref{fig:freespecStag}, resembling the exact overlap operator at $L_s = \infty$.

The structural similarity to the the Pauli-Villars preconditioner is striking. In this case, we could have also started with a pair of anti-Hermitian indefinite operators, ``$i \Gamma_5 D_{PV}$'' for the Pauli-Villars operator and ``$i \Gamma_5 D_{DW}(m)$'' for the domain wall operator. We can now formulate the Pauli-Villars preconditioned domain wall operator as
\be
D_{DW}(m) \rightarrow D_{PV}^{-1} D_{DW}(m) \equiv (i \Gamma_5 D_{PV})^{-1} (i \Gamma_5 D_{DW}(m)),
\ee
where we've made explicit that this preconditioning takes an imaginary indefinite spectrum to a unitary circle, identical in form to
Fig.\ref{sec:specmap}. Beyond
the practical consequence of this observation, it is intriguing to ask
why this is the case. It may hint of a unifying principle for our multigrid algorithm 
common to all three major fermion discretizations in the chiral limit, which
is worthy of additional investigation.

\subsection{Outline of our Three Step Multigrid Implementation}
\label{sec:ThreeSteps}

\paragraph{Pauli-Villars Preconditioning:} 
For the {\bf first step} we need to consider
the ideal Pauli-Villars preconditioner. It is worth re-emphasizing the challenge and importance of  preconditioning the domain wall operator. The domain wall operator on  a $d+1$ dimensional lattice  increases the number of eigenvalues from $N_{ev} = 2^{d/2} \times N_c \times  L^d $ by a factor
of $L_s$. The Pauli-Villars inverse spectra transform is an ideal preconditioner, putting the "bulk" $N_{ev} \times (L_s -1)$ eigenvalues exactly at the cut-off  $1/a$ in physical units. 

Due to the Pauli-Villars operator having a maximally indefinite spectrum, it is most optimally solved via the normal operator. Although the Pauli-Villars normal operator is extremely well conditioned with a positive real spectrum starting at $1/a^2$, its use as a preconditioner is still prohibitively expensive. Instead we consider an approximate Pauli-Villars preconditioner, 
\be
D^{-1}_{PV} =  D^\dag_{PV} [ D_{PV} D^\dag_{PV}]^{-1}  \simeq  D^\dag_{PV}
\label{eq:NormalExp}
\ee
as motivated by low-order expansion of the Pauli-Villars normal operator given in Eq.~\ref{eq:loworderpv}. This approximate operator importantly preserves the property that the spectrum is confined to the right complex half-plane,
\begin{align}
\left(D_{PV}^\dagger D_{PV}\right) D_{PV}^{-1} D_{DW} \ket{\lambda} &= r e^{i\theta} \ket{\lambda}, \ ;
\end{align}
with $-\pi/2 < \theta < \pi/2$ for all eigenvalues $\lambda = r \exp[i \theta] $. 
This  is proven in Appendix~\ref{app:RHPproof} based 
on the positive definite spectra of the normal operator factor and the right-half plane spectrum of $D_{PV}^{-1} D_{DW}$. {\bf This does imply Krylov solvers such as BiCGStab can be directly applied to this approximate operator.}

\begin{figure}[t]
    \centering
    \includegraphics[width = 0.9\textwidth]{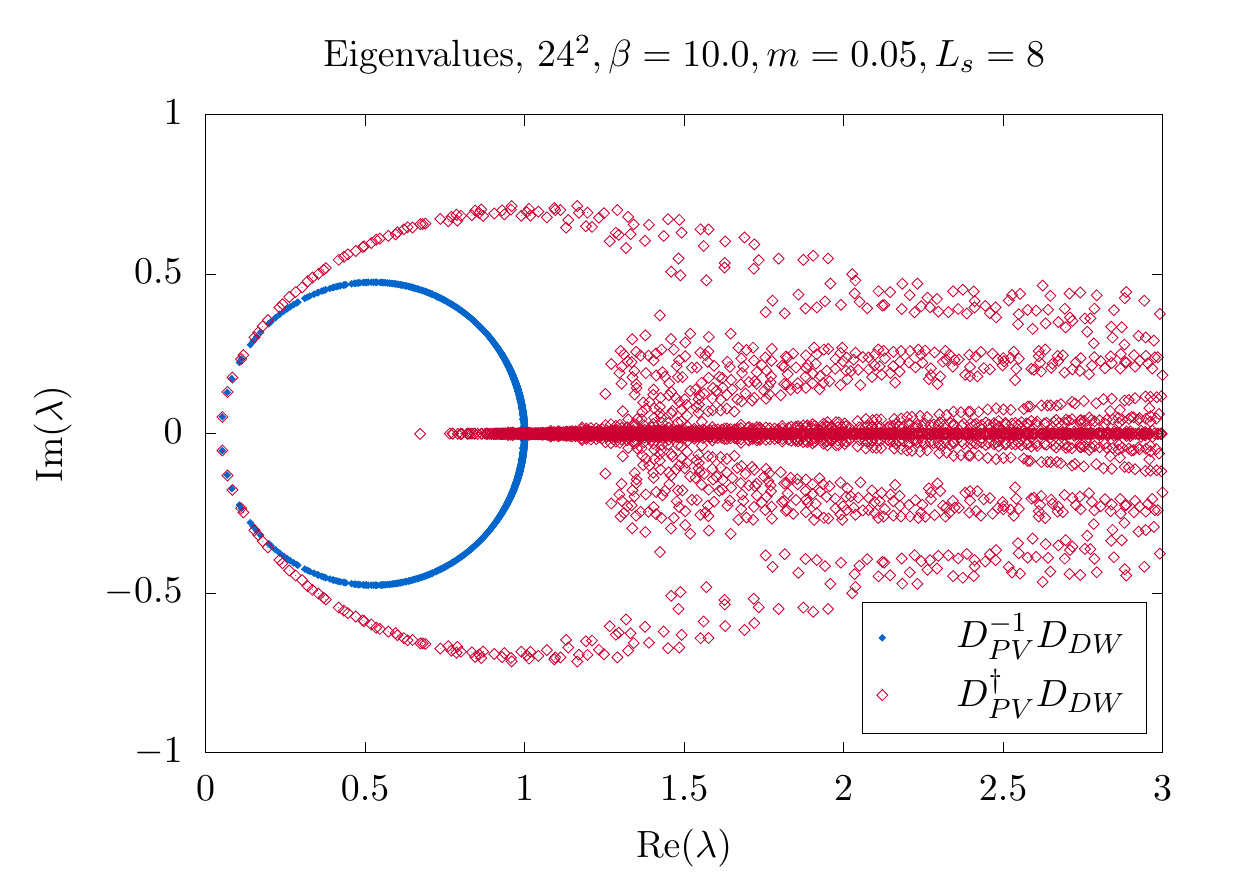}
    \caption{The spectrum of our target multigrid operator, $D_{PV}^\dag D_{DW}$, compared with the effective overlap spectrum, $D_{PV}^{-1} D_{DW}$. For clarity of presentation we truncate the x-axis; the spectrum of $D_{PV}^\dag D_{DW}$ extends out to Re($\lambda$)$\approx 25$.}
    \label{fig:overlap_and_approx}
\end{figure}

While Fig.~\ref{fig:overlap_and_approx} is consistent with this property, the qualitatively strong match between the low eigenvalues of the two operators suggests we can make a much stronger statement. Indeed the two operators are nearly identical, with deviations confined to larger eigenvalues in the approach to the cut-off scale $\pi/a$. This is again motivated by the free field limit where we prove for $M_5 = -1$ 
\be
 D_{PV} D^\dag_{PV} = 1  + \mathcal{O}(p^4)\;,
\ee
and as a result the spectrum of the approximation is valid up to $\mathcal{O}(p^4)$ corrections.
Indeed in the free theory, the additive operator to  $1$ (or $1/a^2$ in physical units) is  positive definite for all momenta. Further details on this can be found in Appendix~\ref{app:lowepequiv}.

We note that a point of future investigation could be approximating $(D_{PV} D_{PV}^\dagger)^{-1}$ by a low-order polynomial in the normal operator as opposed to truncating it to $1$. Given the success of the truncation to 1, it is unclear if higher order polynomials would be worth the additional computational burden.

\paragraph{Wilson Kernel MG Projection:}
In the {\bf second step}, we introduce a
coarsening projection using the familiar Galerkin projecting developed  for Wilson MG acting independently on the Wilson kernel for each of the $s = 1,2,\cdots L_s$ slices,
\be
\widehat D^{W}_{\widehat{x},\widehat{x}'} = \Pmg_{\widehat{x},x}^\dag D^W_{x,x'}  \Pmg_{x',\widehat{x}'} \quad \mbox{or} \quad \widehat D_{W} = \Pmg^\dag D_W  \Pmg \; .
\ee
Here color and spin indices are implicit, and on the right we have further followed the convention of Wilson MG  by suppressing the indices of the d-dimensional space-time lattice. The projection preserves $\gamma_5$ so 
that $\gamma_5 \mathbb{P} =  \mathbb{P} \sigma_3$ and
\be
\Pmg^\dag (1 \pm \gamma_5)  \Pmg =  1 \pm \sigma_3.
\label{eq:gamma5}
\ee
With the normalization convention of the restrictor, $\Pmg^\dag \Pmg = {\mathbb I}$, giving the identity
operator on the coarse vector space {\bf and}
the fact that $ D_{DW}(m)$ is a {\bf linear functional}
of the Wilson kernel, we have
\be
\widehat D_{W} = \Pmg^\dag D_W  \Pmg \implies \widehat D_{DW}(m) = \Pmg^\dag D_{DW}(m) \Pmg.
\ee
with the implicit redefining on the right of the
restrictor as diagonal in  s-space: $\Pmg \rightarrow \Pmg \delta_{s,s'} $. This notational {\em slight of hand} is common practice in the physics
literature with tensor expressions. For example in
Eq.~\ref{eq:gamma5} we also implicitly
redefined  $\gamma_5$  as diagonal in  color and d-dimensional space-time.

Of course this  factorization  also applies  to the Pauli-Villars term
and to generalized domain wall formulations such as M\"obius, Zolotarev etc. However this factorization does {\bf not}
apply to non-linear functional of the kernel such as the preconditioned product
\be
(\Pmg^\dag D_{PV}\Pmg^\dag)^{-1}  \Pmg^\dag D_{DW}(m)\Pmg^\dag \ne \Pmg^\dag D^{-1}_{PV} D_{DW}(m)\Pmg^\dag
\label{eq:KernelProject}
\ee
or the overlap operator. For these the  kernel  projection  does not commute
with the operator. To make this clear we explicitly write the
coarsened form of the domain wall and Pauli-Villars operators,
\begin{align}
\widehat{D}_{DW}\left(m\right) &= \left[\begin{array}{ccccc}
\widehat{D}_W(M_5) + 1 & \widehat{P}_{-} & 0 & \cdots & -m\widehat{P}_{+} \\
\widehat{P}_{+} & \widehat{D}_W(M_5)+1 & \widehat{P}_{-} & \cdots & 0 \\
0 & \widehat{P}_{+} & \widehat{D}_W(M_5)+1 & \cdots & \vdots \\
\vdots & \vdots & \vdots & \ddots & \widehat{P}_{-} \\
-m\widehat{P}_{-} & 0 & \cdots & \widehat{P}_{+} & \widehat{D}_W(M_5)+1
\end{array}\right],
\label{eq:coarsenDW}
\end{align}
where $\widehat{P}_{\pm} = \frac{1}{2}\left(1 \pm \sigma_3\right)$. This coarsened operator has  identical algebraic structure as original fine-level Domain wall operator. So when applying
a coarse Pauli-Villars preconditioner $\widehat D^{-1}_{PV}$, the exact same algebraic
manipulations for the \BNO factorization carry over. Again the bulk modes are
moved to the cut-off and ``chiral'' modes are  confined to the boundary to 
form an effective coarse $\widehat D_{ov}(m)$ operator. This recursive construction is
the key element for our DW multigrid construction.

\paragraph{Truncated projection/prolongation:} 

Lastly, in the {\bf third step}, we define a convention for residual coarsening and error correction prolongation. One straightforward approach would be to coarsen and prolongate across all $L_s$ slices. Instead, we find it is possible and in fact advantageous to only restrict and prolong the boundary contribution for the residual coarsening and the error correction, respectively. We assert this convention here and use it in general going forward, however we quantitatively study only acting on the boundary as opposed to the entire bulk in Sec.~\ref{sec:transferop}.

In the convention where the boundary is at the $s = 1$ slice, we define
the projection of residual  and
prolongation of the error on the boundary by
\be
\widehat{r}_s =   
  \begin{cases}
   \quad  \Pmg^\dag~r_1  & \text{for}\quad  s = 1 \\
   \quad 0 &    \text{for} \quad  s > 1
  \end{cases}
  %\label{eq:TruncProject}
  %\ee
  \quad \mbox{and} \quad 
  %\be
  e_s = 
   \begin{cases}
   \quad  \Pmg~\widehat e_1  & \text{for}\quad  s = 1 \\
   \quad 0 &    \text{for} \quad  s > 1
  \end{cases} \; .
  \label{eq:TruncProjectInterpolate}
  \ee
  The $s > 1$ elements  are  restored  efficiently by the outer solver and as a smoother correction.
  This is inspired by the {\it Sec 3.2 Overlap to bulk Domain Wall reconstruction} procedure in BNO~\cite{Brower:2012vk}. It may be surprising, but this is still effective despite the fact we are using $\widehat{D}_{PV}^\dagger$ in place of $\widehat{D}_{PV}^{-1}$. We discuss this further below, but in brief, this can be motivated as reasonable again by equivalence,  $D_{PV}^\dag \sim D_{PV}^{-1}$, 
  up to the quadratic order in momenta in the free theory.

All of these elements come with a variety of potential 
parameters for optimization on each level. For example since we are only transferring the residual and error correction on a single slice
we are also able to reduce the extra dimension on the 
coarse levels. We will denote $L_s$ for the first coarsening as $\widehat{L}_s$, on the second coarsening $\widehat{\widehat{L\, }}_s$, and so on. We will use a reduced $\widehat{L}_s$ and $\widehat{\widehat{L\,}}_s$ going forward, however we quantitatively study varying the coarser $L_s$ in Sec.~\ref{sec:redcoarse}.

\paragraph{Summary:}
The full MG algorithm for the linear system $D_{DW}~x  = b$, formulated as an extension to a K-cycle (i.e., a multigrid cycle where each coarse solve is wrapped in a Krylov solver) is as follows:
\begin{itemize}
    \item Left precondition the system by $D_{PV}^\dag$, giving the
      new linear system
\be
 D_{PV}^\dag D_{DW}~x = D_{PV}^\dag~b
\ee
    \item Perform an MG-preconditioned iterative solve (via GCR, FGMRES, etc.) 
    using the operator $D_{PV}^\dag D_{DW}$.
    \begin{itemize}
        \item Relax on the current residual with $D_{PV}^\dag D_{DW}$,
           known as the pre-smoother.
\item Go the next level with the projected Wilson kernel
\be
  \widehat D_{W}(U, M_5) = \mathbb{P}^\dag D_{W}(U, M_5)
  \mathbb{P}
\ee
to define coarse level  operators, $\widehat D_{DW}$ and  $\widehat D_{PV}$, using Eq.~\ref{eq:coarsenDW}. 
\begin{itemize}
\item Project the residual on the wall using Eq.~\ref{eq:TruncProjectInterpolate}
\item Using a Krylov solver, approximately solve the coarse level system:
\be
 \widehat{D}_{PV}^\dagger \widehat{D}_{DW}~ \widehat{e} = \widehat{r}
\ee
\item 
Prolong the error  with Eq.~\ref{eq:TruncProjectInterpolate} and
correct the solution: $x = x + e$
\end{itemize}
        \item Post-smooth on the accumulated error from the previous
          two steps  with  $D_{PV}^\dag D_{DW}$.
    \end{itemize}
    \item Repeat until the desired tolerance on $||b - D_{DW}~x ||$.
\end{itemize}

Needless to say, there are several knobs to tune, even as far as MG algorithms go. We have explored a few of  these parameters in a preliminary
form in the \ndim{2} two-flavor Schwinger model and have left
unexplored further until testing for \ndim{4} domain
wall methods discussed briefly in the conclusion.

\section{Numerical Tests with the \ndim{2} Schwinger Model}
\label{sec:Results}

We now turn to testing our domain wall MG algorithm on the two-flavor Schwinger model~\cite{PhysRev.128.2425,Smilga:1996pi,Adams:2009eb}.  As with the cases of Wilson~\cite{Brannick:2007ue} and staggered MG~\cite{Brower:2018ymy}, the Schwinger model is a useful framework for the development and testing of algorithms for QCD~\cite{Vranas:1997da}. As a low-dimensional prototype model it has the advantage of enabling the rapid exploration of a wide variety of alternative features in a serial laptop code. This can be used to demonstrate validity of an MG algorithm and guide the subsequent application to QCD and software tuning at scale on modern GPU accelerated systems. The importance of this two-step approach can not be over emphasized. 

\begin{figure}[t] \centering
  \includegraphics[width=0.90\linewidth]{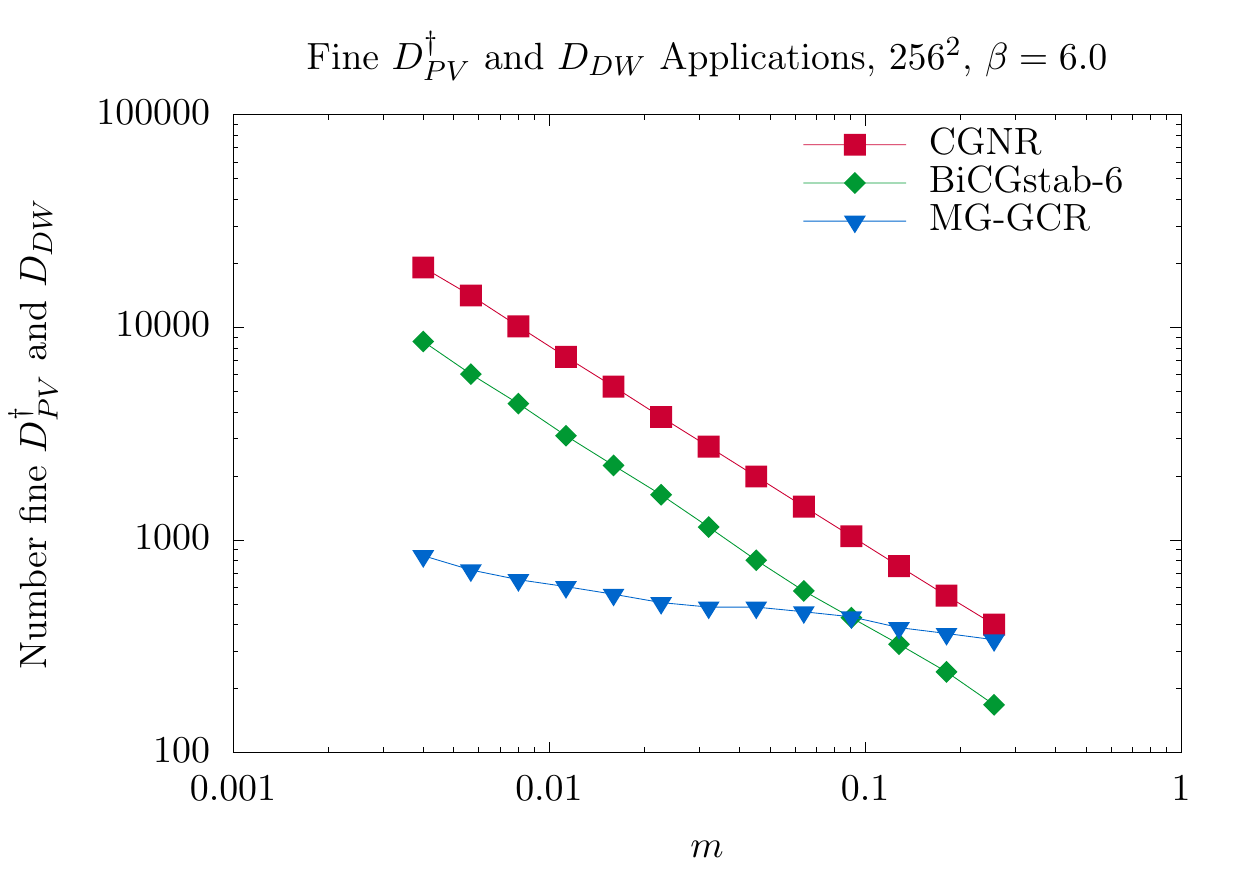}
  \caption{The number of domain wall operator ($D_{DW}, D_{DW}^\dagger, D_{PV}^\dagger$) applications required for a solve to a tolerance of $10^{-10}$ using CGNR on $D_{DW}^\dagger D_{DW}$, BiCGStab-6 on $D_{PV}^\dagger D_{DW}$, and multigrid for a representative fixed $\beta$ and volume. Note that this is a log-log plot.}
\label{fig:iterationscompare}
\end{figure}

For our investigations in the interacting case, we have fixed $M_5 = -1.05$ relative to the correct free-field value $M_5 = -1$ and $L_s = 16$ as a representative large value. Our current performance of  the DW MG algorithm outlined above is illustrated in Fig.~\ref{fig:iterationscompare} and the accompanying Table~\ref{sec:DWformalism}. The one exception to these parameters is our study of the continuum limit, where we explore the addition of deflation on the coarsest level.

Before describing  the details, it is apparent that our basic algorithm vastly improves scaling in the approach to the continuum and chiral limit, nearly eliminating critical slowing down. Our study of additional deflation in the continuum limit, given in Table~\ref{tab:zerospacing}, suggests that the lack of perfect scaling shown in Fig.~\ref{fig:iterationscompare} can be improved with deflation. This is supported by ongoing investigations of deflation of the coarsest level with both twisted mass and HISQ fermions in \ndim{4} QCD and inspiration from~\cite{Yamaguchi:2016kop}.

\begin{table}[tp]
\center
\begin{tabular}{cccc|ccc}
\hline
 $L$ & $m$ & $\beta$ & Defl. & Fine $D_{PV}, D_{DW}$ & Int. avg. iter. & Coarsest avg. iter. \\
\hline
\hline
 64 & 0.01 & 3.0 & N & 820 & 2.88(07) & 2.62(06) $\times 10^3$ \\
\hline
 128 & 0.005 & 12.0 & N & 484 & 2.85(17) & 4.71(42) $\times 10^3$ \\
\hline
 256 & 0.0025 & 48.0 & N & 460 & 3.32(23) & 1.06(11) $\times 10^4$ \\ 
\hline\hline
 64 & 0.01 & 3.0 & Y & 820 & 2.62(09) & 51.2(3.9) \\ 
\hline
 128 & 0.005 & 12.0 & Y & 412 & 2.29(14) & 68.6(0.9) \\ 
\hline
 256 & 0.0025 & 48.0 & Y & 412 & 2.00(09) & 93.2(2.0) \\ 
\hline
\end{tabular}
\caption{\label{tab:zerospacing}The behavior of our MG algorithm in the approach to the continuum limit at constant physics and physical volume. We see an improvement in convergence as we approach the continuum limit, noting that stable convergence depends on an inexpensive deflation of the coarsest level. We chose to deflate 128 eigenvectors on the coarsest level.}
\end{table}

\FloatBarrier

\subsection{Algorithmic Details and Analysis}

\begin{table}[tp]
\center
\begin{tabular}{rll}
\hline
 & parameter & \\
\hline
setup & setup operator & Normal operator, $D_W D_{W}^\dagger$ \\
 & setup solver & CG \\
 & max iterations & 250 \\
 & max residual tolerance per null vector & $10^{-4}$ \\
 & number of null vectors, level 1 ($n_{vec}^1$) & 8 \\
 & size of aggregate block, level 1 & $4^2$ \\
 & number of null vectors, level $\ell > 1$ ($n_{vec}^\ell$) & 12 \\
 & size of aggregate block, level $\ell > 1$ & $2^2$ \\
 & number of levels $\ell_{\mbox{\scriptsize max}}$ & 3 \\
\hline
 solver, level 1 & operator & $D_{PV}^\dagger D_{DW}$ \\
 & restart length of GCR & 16 \\
 & relative residual tolerance & $10^{-10}$ \\
 & GCR iterations for pre-, post-smooth & 0, 8 \\
\hline
 solver, level 2 & operator & $\widehat{D}_{PV}^\dagger \widehat{D}_{DW}$ \\
 & $\widehat{L}_s$ & 4 \\
 & max iterations & 16 \\
 & restart length of GCR & 16 \\
 & relative residual tolerance & 0.25 \\
 & GCR iterations for pre-, post-smooth & 0, 8 \\
\hline
solver, level 3 & operator & $(\widehat{\widehat{D\, }}{\,}_{PV}^\dagger \widehat{\widehat{D\, }}_{DW})^\dagger (\widehat{\widehat{D\, }}{\,}_{PV}^\dagger \widehat{\widehat{D\, }}_{DW})$ \\
 & $\widehat{\widehat{L\, }}_s$ & 4 \\
 & solver & CGNR \\
 & relative residual tolerance & 0.05\\
 & maximum iterations & 1024\\
\hline
\end{tabular}
\caption{\label{tab:kcycle}Relevant fixed parameters we use for our K-cycle. For consistency, we use the same setup parameters throughout the procedures described in this paper. For the setup we have tuned the Wilson operator to the critical mass.}
\end{table}

A benefit of our MG algorithm is its setup is the same as the setup for the traditional $\gamma_5$-preserving MG algorithm for Wilson fermions. In Eq.~\ref{eq:coarsenDW} we see that the coarsened domain wall operator takes a Galerkin-projected Wilson operator as a kernel. For this reason we expect the setup to be roughly the same cost as one five-dimensional domain wall solve. We have tuned the Wilson operator to the critical mass for near-null vector generation. Near-null vectors are generated by relaxing on the homogeneous normal system with a Gaussian-distributed initial guess followed by a chiral doubling via $\frac{1}{2}\left(1+P_\pm\right)$ to preserve a $\sigma_3$-Hermiticity on the coarser levels~\cite{Brower:2018ymy,Brannick:2007ue}.

For the solve, the main structure of the K-cycle is unchanged relative to our previous work. Fixed parameters related to the setup and the K-cycle (target tolerance on each level, etc) are described in Table~\ref{tab:kcycle}. We have standardized on $L_s = 16$ throughout this investigation, though explorations for other values of $L_s$ are discussed in Sec.~\ref{sec:explore}.

Inspired by the work of~\cite{Yamaguchi:2016kop}, we tested adding a deflation step to the coarsest level, resulting in a four-level algorithm. This is important for addressing critical slowing down on the coarsest level, as has been also seen in \ndim{4} studies for Wilson and staggered fermions. For conciseness of presentation we do not explore this past the results given in Table~\ref{tab:zerospacing}.
    
To study the viability of our MG algorithm, we consider sweeps in input fermion mass $m$ at both fixed $\beta = 6.0$, varying the 2-d lattice volume and at fixed volume $256^2$, varying the bare coupling $\beta$. An ideal MG algorithm shifts critical slowing down to the coarsest level, corresponding to mass-independent behavior on the finest and intermediate levels. Further, it should be insensitive to the volume at constant physics. We are interested in an algorithm that works for all reasonable values of $\beta$, however in practice it only needs to works for large $\beta$ approaching the continuum at fixed physical correlation lengths well below the UV cut-off.

\begin{figure}[t] \centering
  \includegraphics[width=0.45\linewidth]{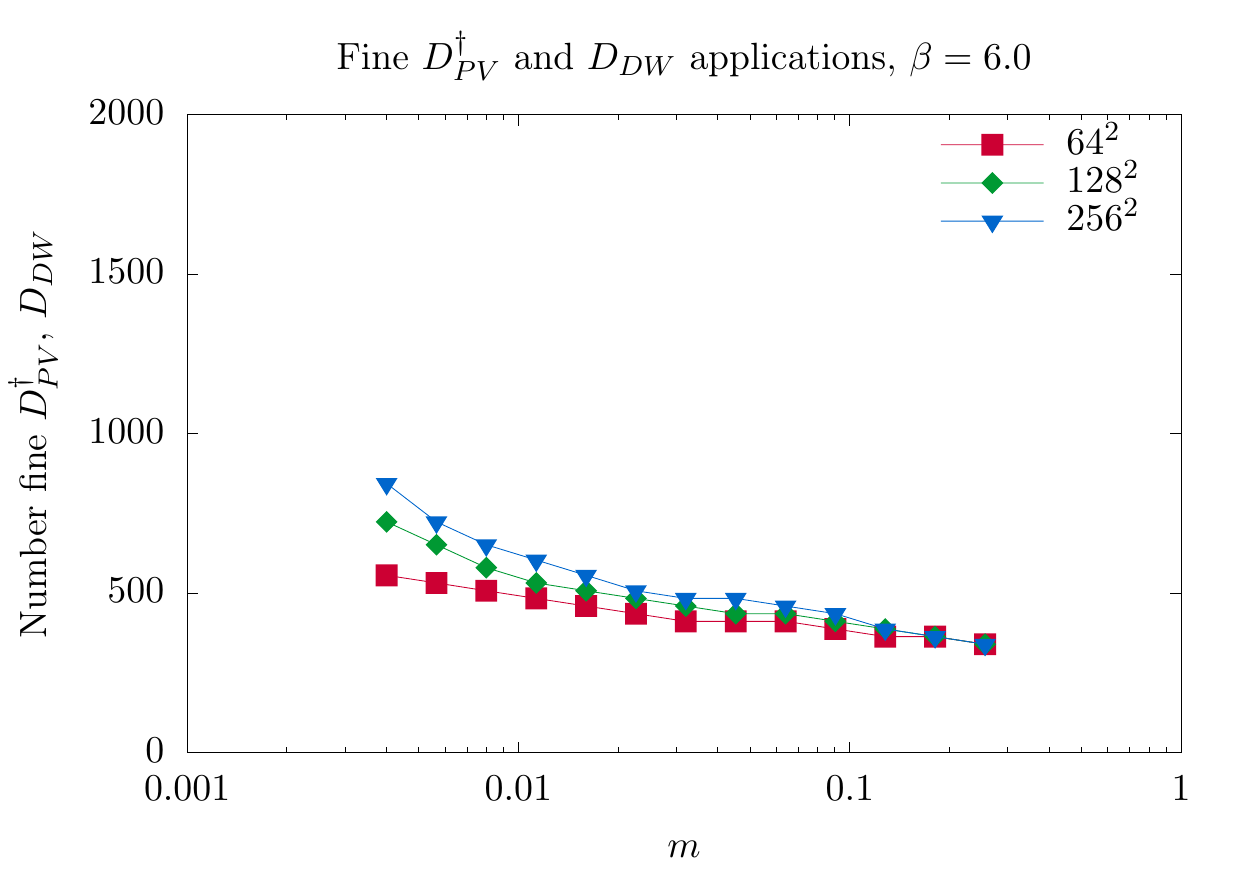}
~\includegraphics[width=0.45\linewidth]{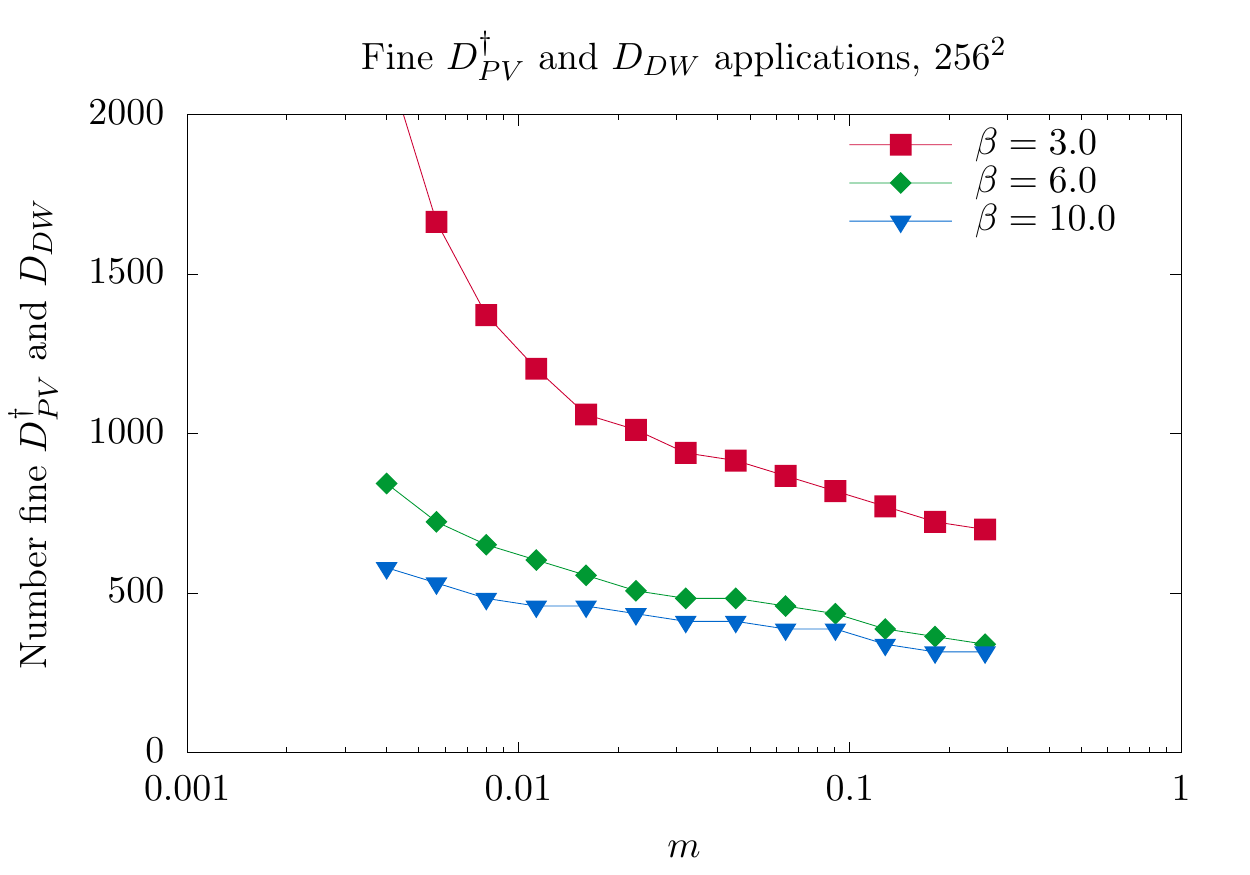}
  \caption{The number of applications of the fine domain wall operator (separately $D_{DW}$ and $D_{PV}^\dagger$ ) within an MG-preconditioned solve as a function of mass. On the left, we consider fixed $\beta = 6.0$, and on the right, fixed volume $256^2$.}
\label{fig:iterationsfine}
\end{figure}

\subsection{Elimination of critical slowing down}

In Fig.~\ref{fig:iterationsfine} we consider the number of fine domain wall operator applications as a function of the input fermion mass, which is proportional to the number of outer GCR iterations. On the left, we see that for fixed $\beta$ critical slowing down has been largely eliminated, albeit there is still a small increase at vanishing  mass. On the right, we see that for a fixed volume, the MG algorithm shows an improved fermion mass independence in the approach to the continuum limit. The algorithm is unsuccessful for our coarsest $\beta = 3.0$, however this corresponds to a gauge correlation length of $\ell_\sigma \approx 2.4$, which is pushing into an unphysical regime. Similar effects have been noticed in staggered MG~\cite{Brower:2018ymy}. As noted in Table~\ref{tab:zerospacing}, preliminary investigations of deflation on the coarsest level lead to a further reduction in iteration count and, by extension, operator application count on the fine level, though not a complete elimination of the increase as a function of decreasing mass.

\paragraph{Intermediate level:}

\begin{figure}[t] \centering
 \includegraphics[width=0.47\linewidth]{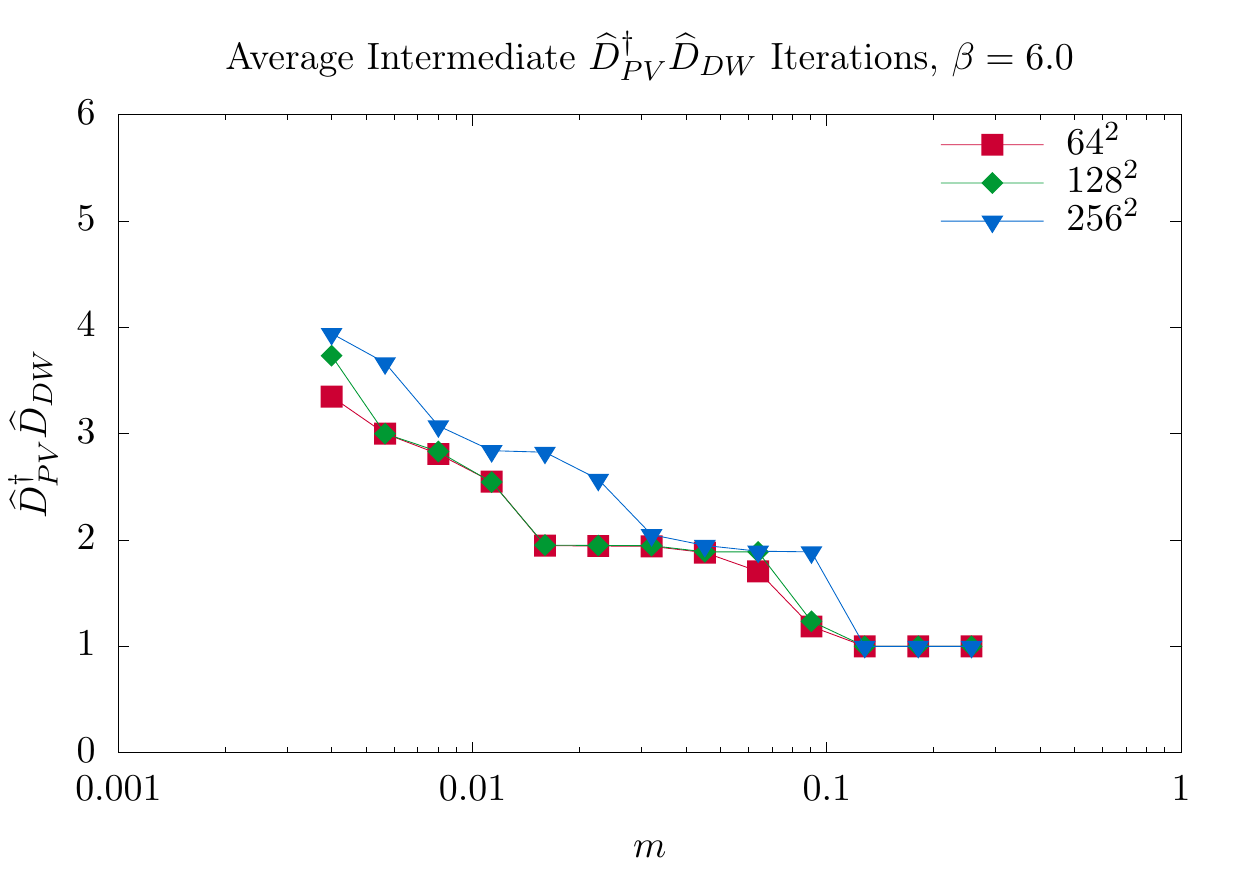}
~\includegraphics[width=0.47\linewidth]{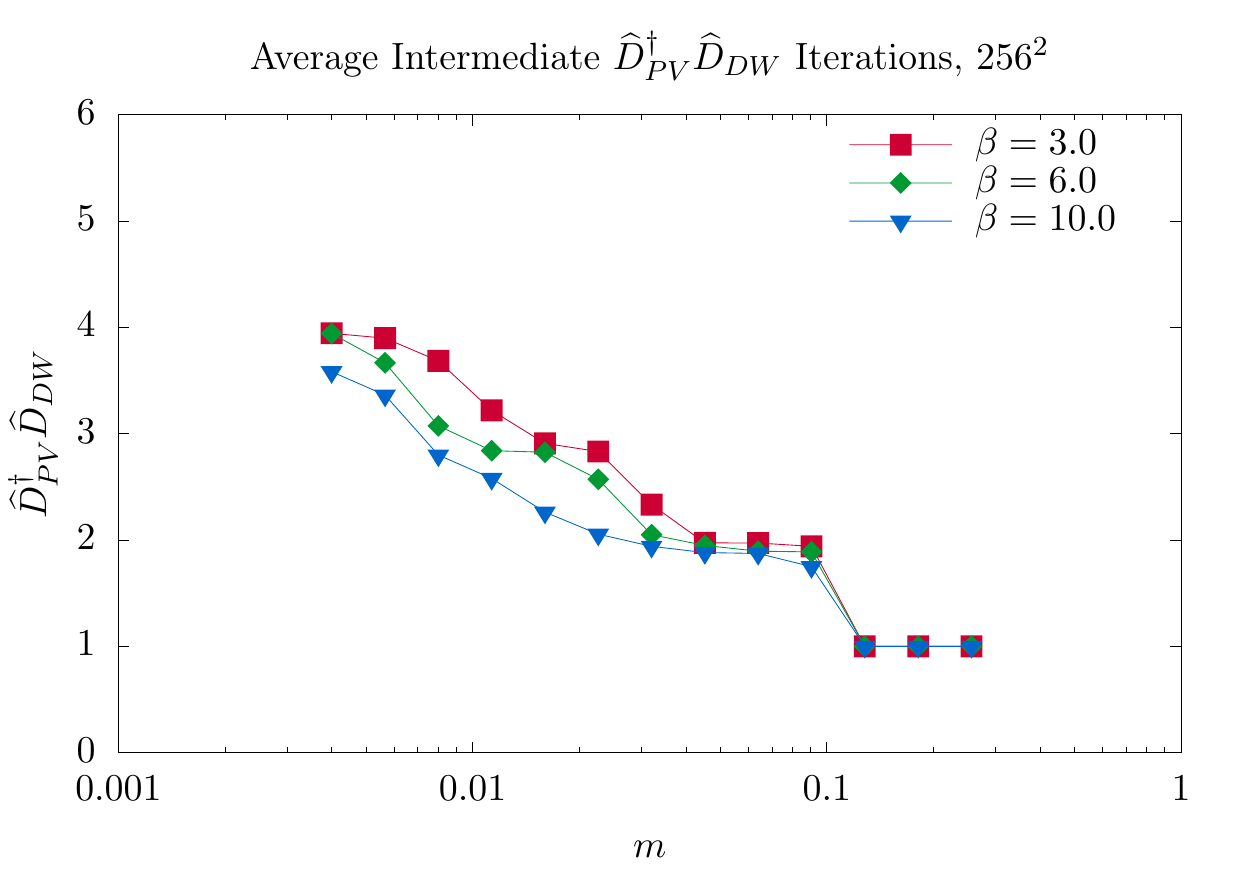}
\caption{The average number of iterations for the inner Krylov solve as a function of mass. On the left, we consider fixed $\beta = 6.0$, and on the right, fixed volume $256^2$. Note that this is a log-linear plot.}
\label{fig:iterationsinter}
\end{figure}

In Fig.~\ref{fig:iterationsinter} we consider the average number of GCR iterations on the intermediate level as a function of the input fermion mass. We see that the average iteration count is roughly independent of the coupling $\beta$ and the volume, which is encouraging. There is a weak mass dependence to the iteration count, however we note that the growth in iteration count appears to be weaker than power law. This is a significant improvement over the power-law dependence that is traditional of critical slowing down. Preliminary investigations of deflation on the coarsest level lead to a reduction in iteration count on the intermediate level, though not a complete elimination of the increase as a function of decreasing mass.

\paragraph{Coarsest level:}

\begin{figure}[t] \centering
  \includegraphics[width=0.47\linewidth]{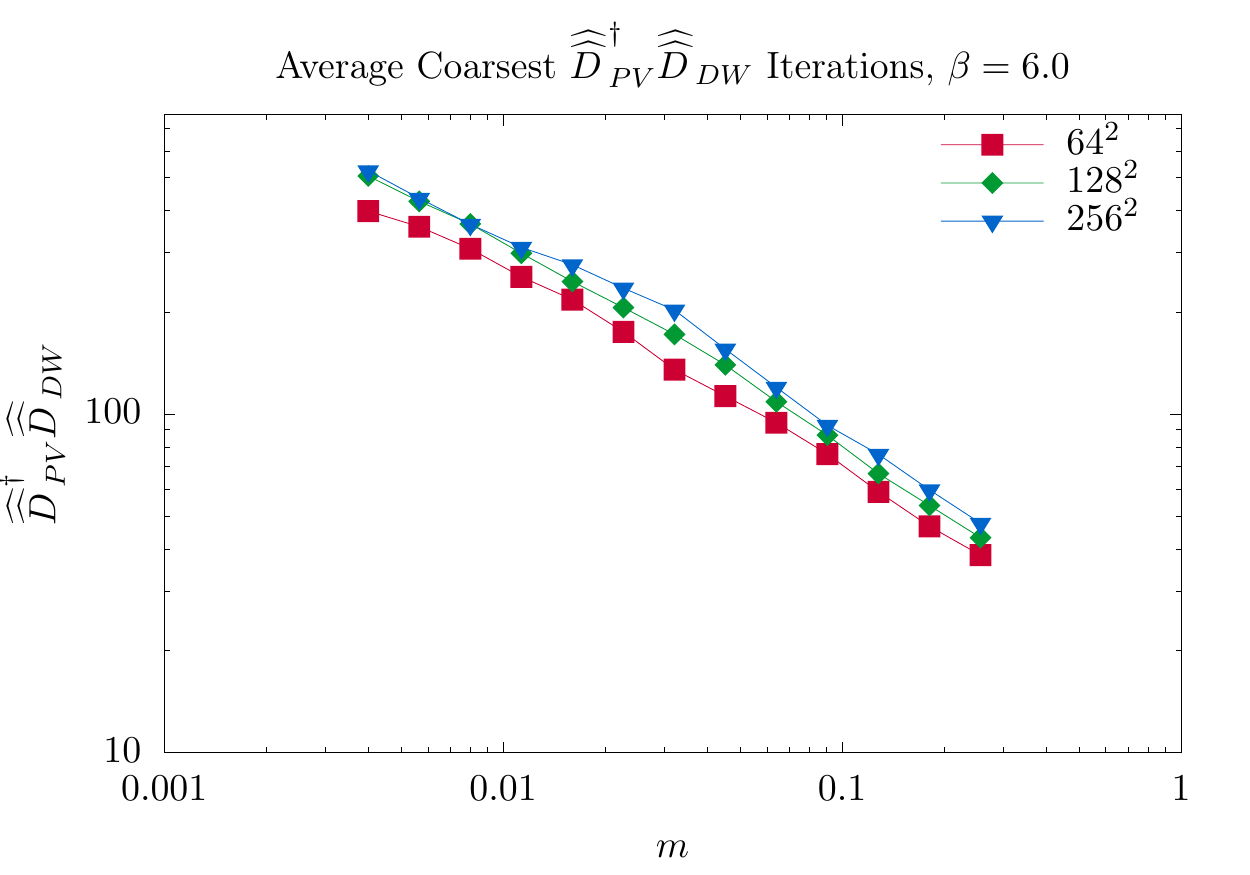}~\includegraphics[width=0.47\linewidth]{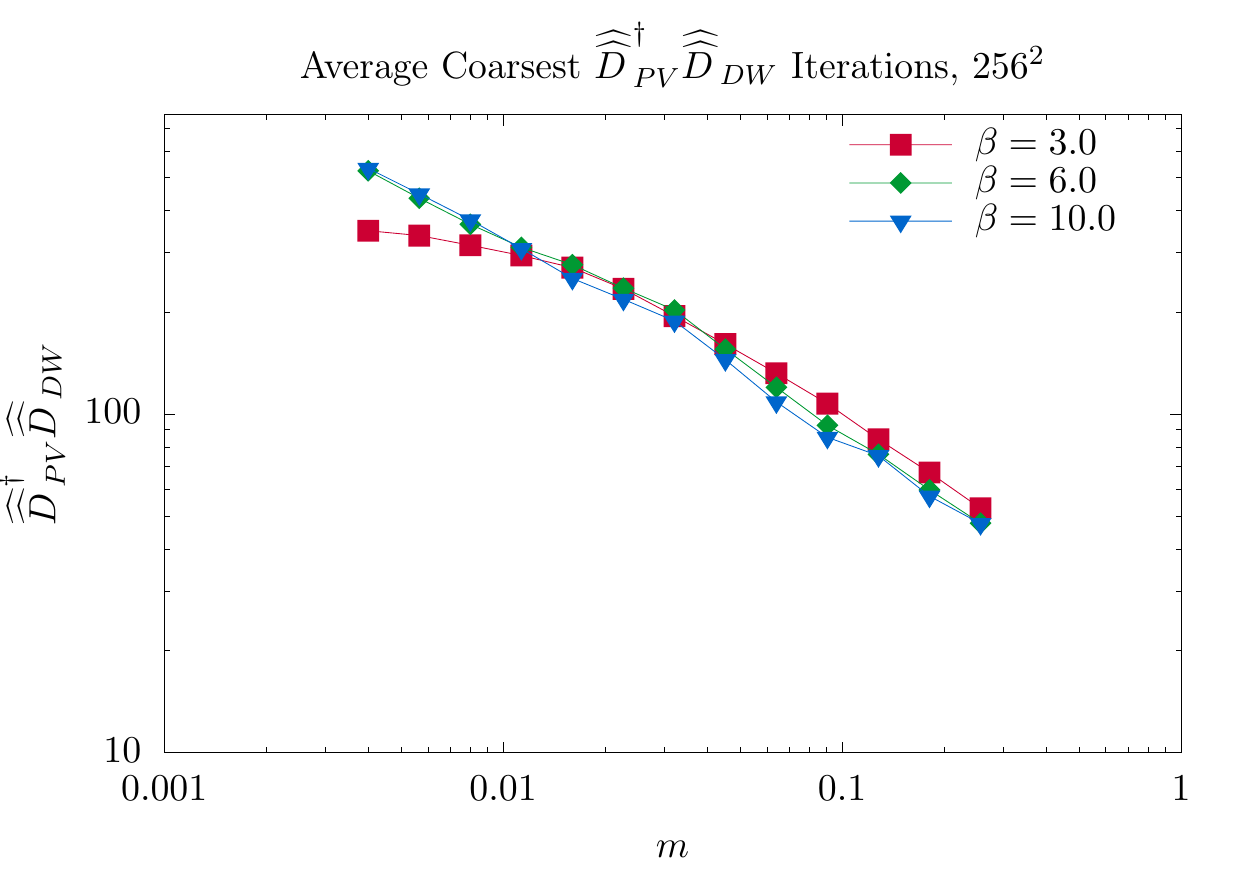}
  \caption{The average number of iterations of CGNR on the coarsest level for, on the left, fixed $\beta$ and on the right, fixed volume. Note that this is a log-log plot.}
\label{fig:iterationscoarsest}
\end{figure}

In the previous two paragraphs we have demonstrated the elimination of critical slowing down from the finest and the intermediate level. This is because critical slowing down has been shifted to the coarsest level. In Fig.~\ref{fig:iterationscoarsest} we consider the average number of CGNR iterations on the coarsest level as a function of the input fermion mass. In contrast to plots for the fine and coarse levels, here we present this data on a log-log plot to examine power behavior. In both the left and right panels, we see behavior consistent with power-law divergence of the iteration count independent of volume and $\beta$, showing that critical slowing down has been successfully shifted to the coarsest level. As has been seen in studies with twisted clover and HISQ fermions in \ndim{4}, this final critical slowing down can be efficiently eliminated by deflation.

\paragraph{Comparison with direct solve:}
In the previous paragraphs we have demonstrated a MG algorithm which shifts critical slowing down from the finest level down to the coarsest level. In Fig.~\ref{fig:iterationscompare}, we can see a stark contrast in behavior between our MG algorithm and CGNR directly on the domain wall operator. While the number of fine domain wall operator applications scales with only weak mass dependence in the case of our MG algorithm, there is a strong power-law dependence present for CGNR. This is the critical slowing down which has shifted to the coarsest level in our MG algorithm.

We also considered applying the BiCGStab-$\ell$ Krylov solver directly to $D_{PV}^\dagger D_{DW}$. We can do this because, in contrast to $D_{DW}$ in isolation, $D_{PV}^\dagger D_{DW}$ obeys the half-plane condition as proven in Appendix~\ref{app:RHPproof}. While this approach unsurprisingly still demonstrates critical slowing down, there is a marked reduction in fine operator applications. This could be of immediate use for computing domain wall propagators for four-dimensional QCD.

\section{Discussion}
\label{sec:explore}

The MG algorithm described above, while generally successful, does introduce several
components that are worth understanding better and very likely can lead to further improve
performance. Even if we were to use the full effective overlap operator, $D_{PV}^{-1} D_{DW}$, the assumption that we need only prolong and restrict the boundary mode when going between levels is non-trivial. That this continues to be adequate with our approximation, $D_{PV}^\dag D_{DW}$ is even more surprising. Also 
the benefit of reducing $L_s$ on the coarsened levels begs a better understanding. Here is  our initial effort to explore these issues.

As we did in our prior paper on the staggered multigrid~\cite{Brower:2018ymy}, we begin by studying the spectrum of our approximate operator and its ``coarsened'' version to glean some intuition. While we find a strong overlap of physical low eigenvalues between the fine and the coarsened operator, we also find that the coarsened operator includes additional spurious small eigenvalues. Unlike the na\"ive formulation of MG for staggered fermions, these modes do not appear to undermine the success of the algorithm. To probe this phenomena, we consider the local colinearity and the oblique projector as monitors for the quality of our MG preconditioner. We see that transferring only the boundary component of the fine residual and the coarse error correction is essential for the  success of  our  multigrid algorithm, and present a physical argument for why this ``cures'' the problem introduced by the spurious small eigenvalues. In the free field limit presented in detail in Appendix~\ref{app:lowepequiv}, we note that the low momentum modes to order $\mathcal{O}(p^2)$ are fixed
for any $L_s \ge 2$, which maybe an indication of the underlying mechanism. We investigated reducing $L_s$ on the coarser levels, finding that this reduction leads to an improved algorithm relative to using the fine $L_s$ on the coarsened levels.

\subsection{Boundary-Only Transfer Operator}
\label{sec:transferop}

\begin{figure}[t] \centering
  \includegraphics[width=\linewidth]{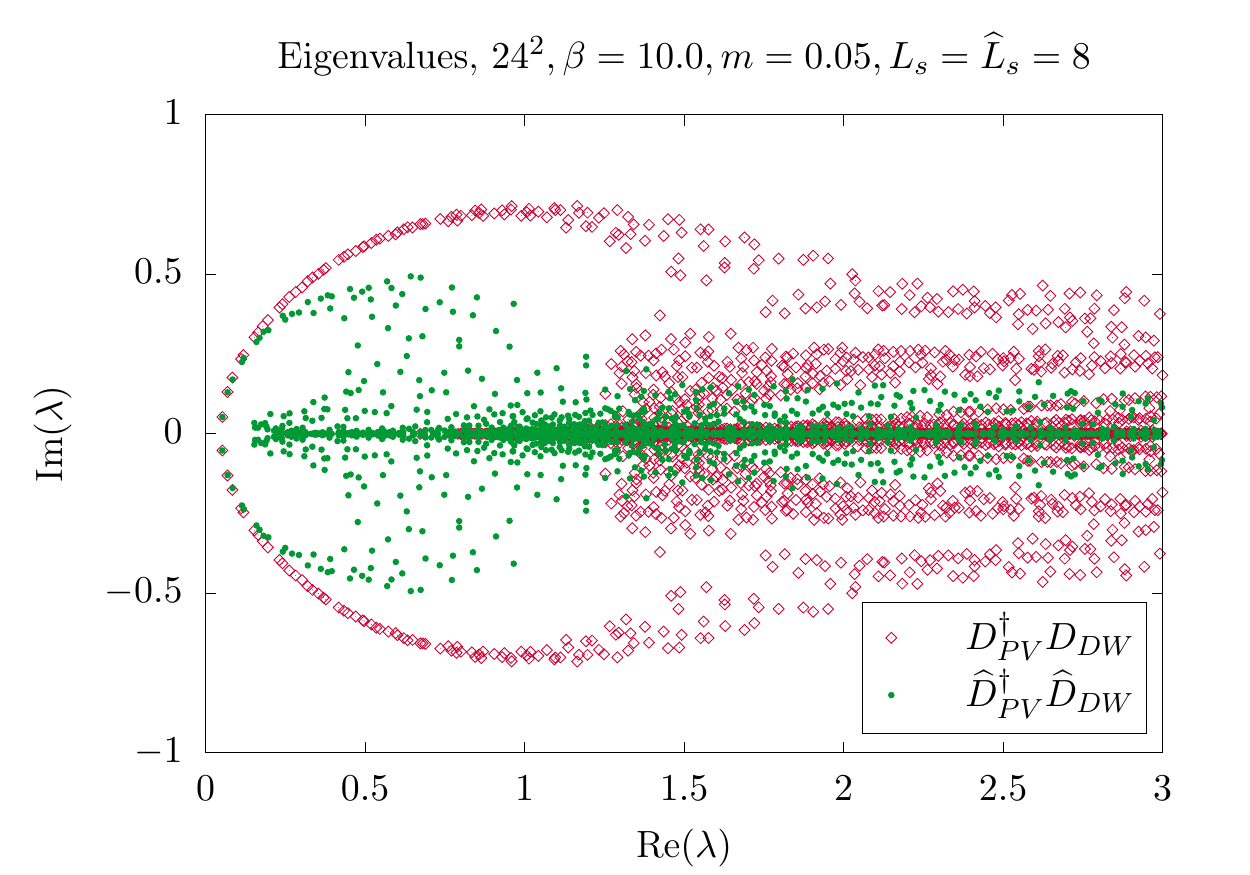}
  \caption{A representative spectrum of $D_{PV}^\dagger D_{DW}$ and its coarsened version $\widehat{D}_{PV}^\dagger \widehat{D}_{DW}$ at fixed $L_{s} = \widehat{L}_s = 8$.}
\label{fig:spectrumfixls}
\end{figure}

In Fig.~\ref{fig:spectrumfixls}, we see that the spectrum of the coarsened operator $\widehat{D}_{PV}^\dagger \widehat{D}_{DW}$ relative to the spectra for the fine operator $D_{PV}^\dagger D_{DW}$ introduces a large number of nearly real low modes. The problem resembles the spurious small eigenvalues that plagued the direct application of Galerkin projection to the staggered operator prior to the K\"ahler-Dirac preconditioning.

In this instance we posit that the saving grace for the domain wall operator is that the low modes of $D_{PV}^{-1} D_{DW}$ are bound to the chiral walls, while higher modes bleed more dominantly into the bulk as suggested by Eq.~\ref{eq:Kinvmatrix}. Based on this we posit that projecting only the boundary modes between levels acts as a filter against the bulk spurious modes.

In Sec.~\ref{sec:ThreeSteps} we noted two ways to formulate the transfer operator between levels. The method we utilize across this paper is only prolonging and restricting the boundary mode as given in Eq.~\ref{eq:TruncProjectInterpolate}. Another formulation would be to repeat the \ndim{2} prolongator/restrictor across the bulk dimension.

As a quantitative approach to this study we follow the investigations of the staggered MG paper~\cite{Brower:2018ymy}. Consider the normalized right eigenpairs of the fine operator, $(\lambda, v_{\lambda})$. Given these we inspect both the \emph{local colinearity}, a measure of preserving the low eigenspace in the least-squares sense, defined by
\be
||(1 - P R)v_\lambda||_2 \; ,
\ee
and the \emph{oblique projector}, defined by
\be
||(1 - P(\widehat{D}^\dagger_{PV} \widehat{D}_{DW})^{-1} R D^\dagger_{PV} D_{DW}) v_\lambda||_2\; ,
\ee
which quantifies the reduction or enhancement of a given error component for a magnitude less than or greater than one. While an error enhancement is not inherently a problem, a very large error enhancement requires a prohibitively expensive compensation at the smoother step.
\begin{figure}[t] \centering
\large{$24^2, \beta = 10.0, m = 0.05, L_s = 8$} \\
 \includegraphics[width=0.47\linewidth]{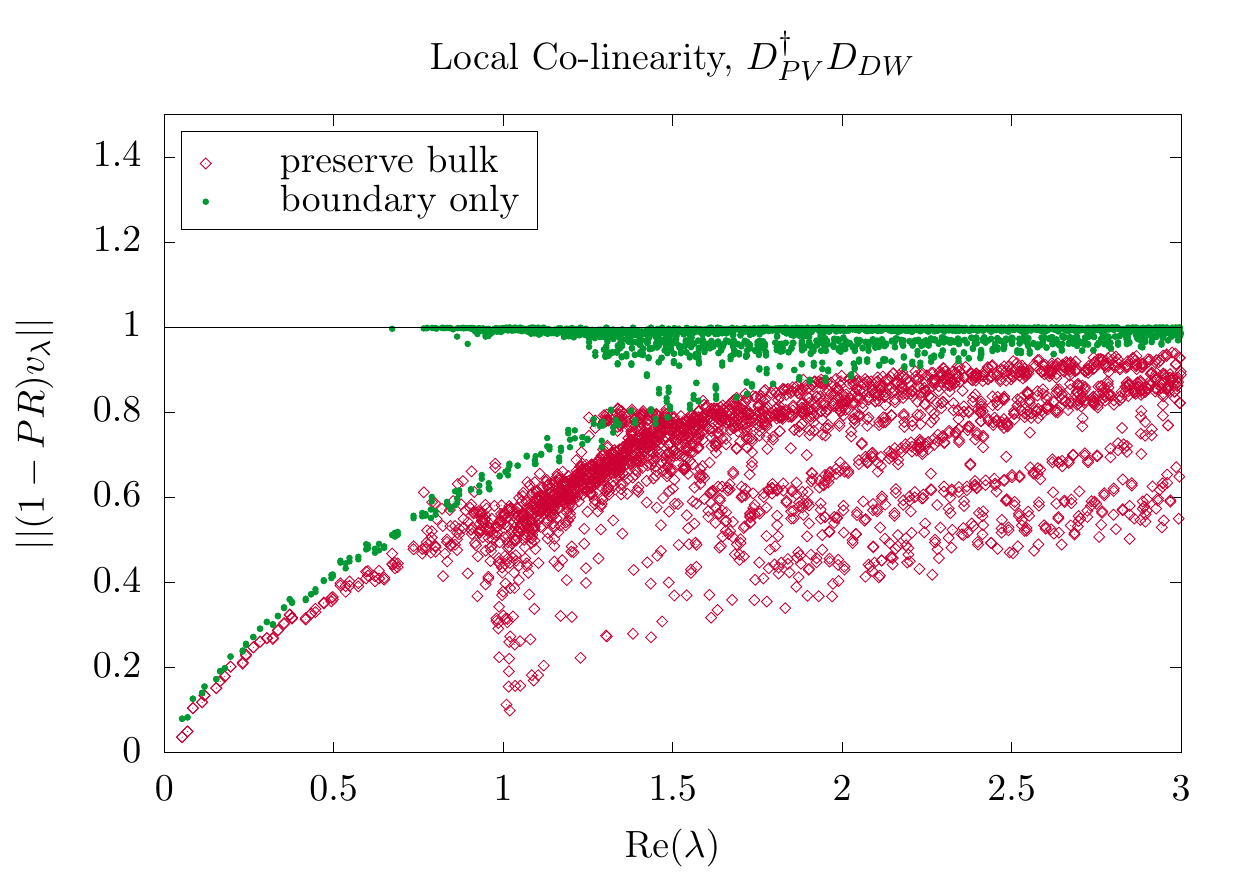}
~\includegraphics[width=0.47\linewidth]{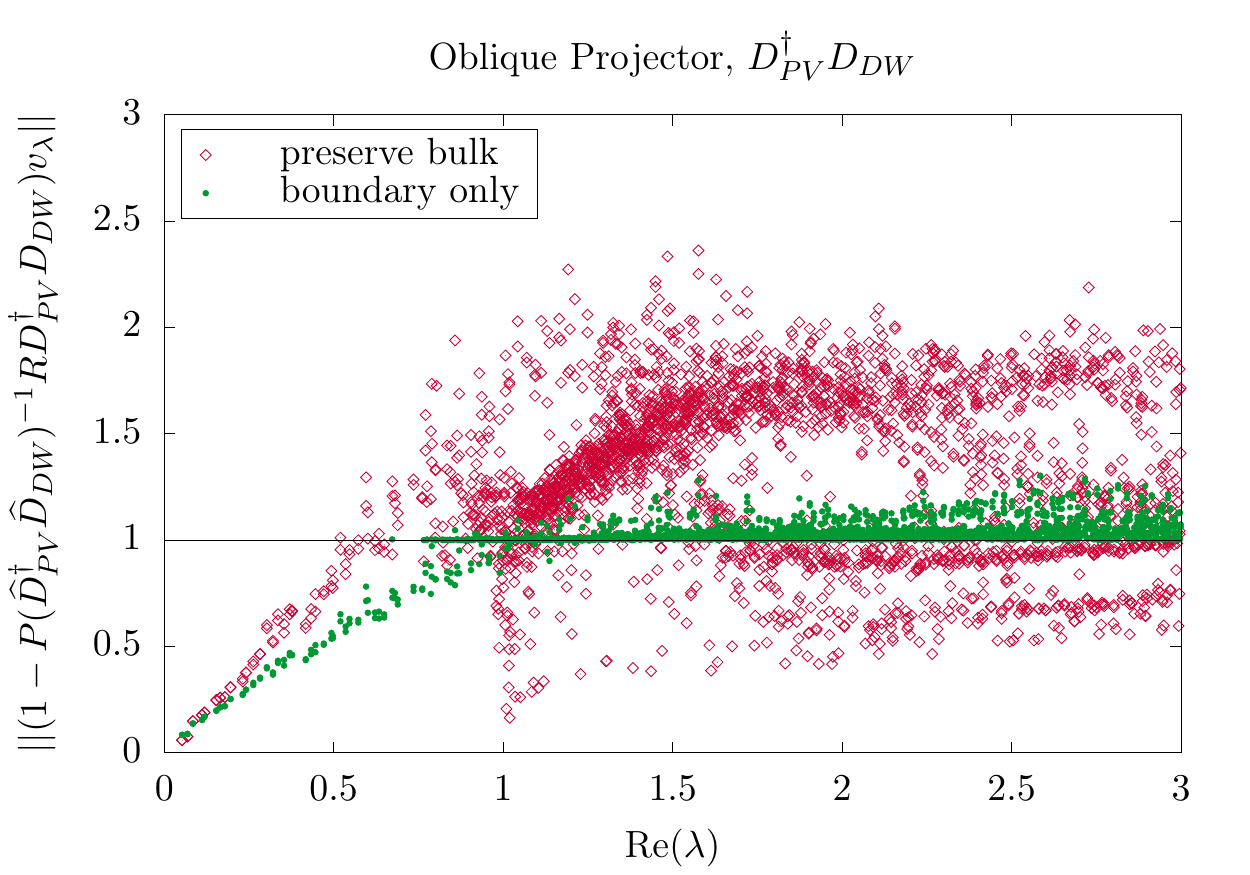}
\caption{On the left, a measurement of local colinearity, and on the right, the effect of
the oblique projector on eigenvectors. Both figures compare a prolongator and restrictor which transfer the boundary and bulk vs just the boundary. The parameters for the MG aggregation are given in Table~\ref{tab:kcycle}.}
\label{fig:bulkboundaryproj}
\end{figure}

In Fig.~\ref{fig:bulkboundaryproj}, we consider the local colinearity and the oblique projector for a given configuration. On the left-hand side, we see that the local colinearity is smaller in magnitude when prolongating and restricting the entire boundary and bulk compared with just transferring the boundary. This is not surprising as the boundary-only transfer operator by construction has a smaller span than the full transfer operator. 
On the other hand, the oblique projector using the boundary-only transfer operator is generally smaller in magnitude than the full transfer operator. This is a good indicator that the coarsening prescription given in the previous section combined with a transfer operator acting only on the boundary modes leads to an improved  MG algorithm.

\begin{figure}[ht] \centering
\large{$24^2, \beta = 10.0, m = 0.05, L_s = 8$} \\
 \includegraphics[width=0.47\linewidth]{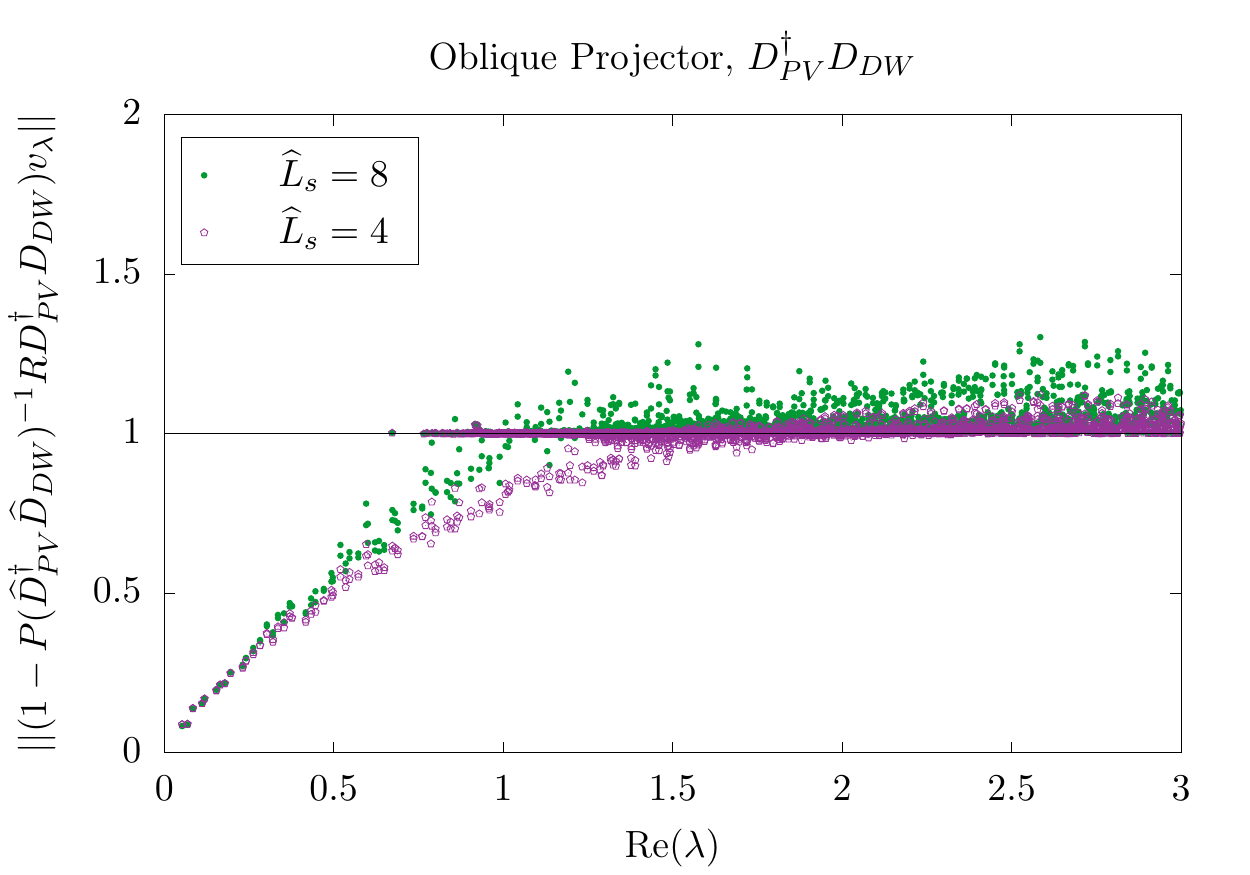}
~\includegraphics[width=0.47\linewidth]{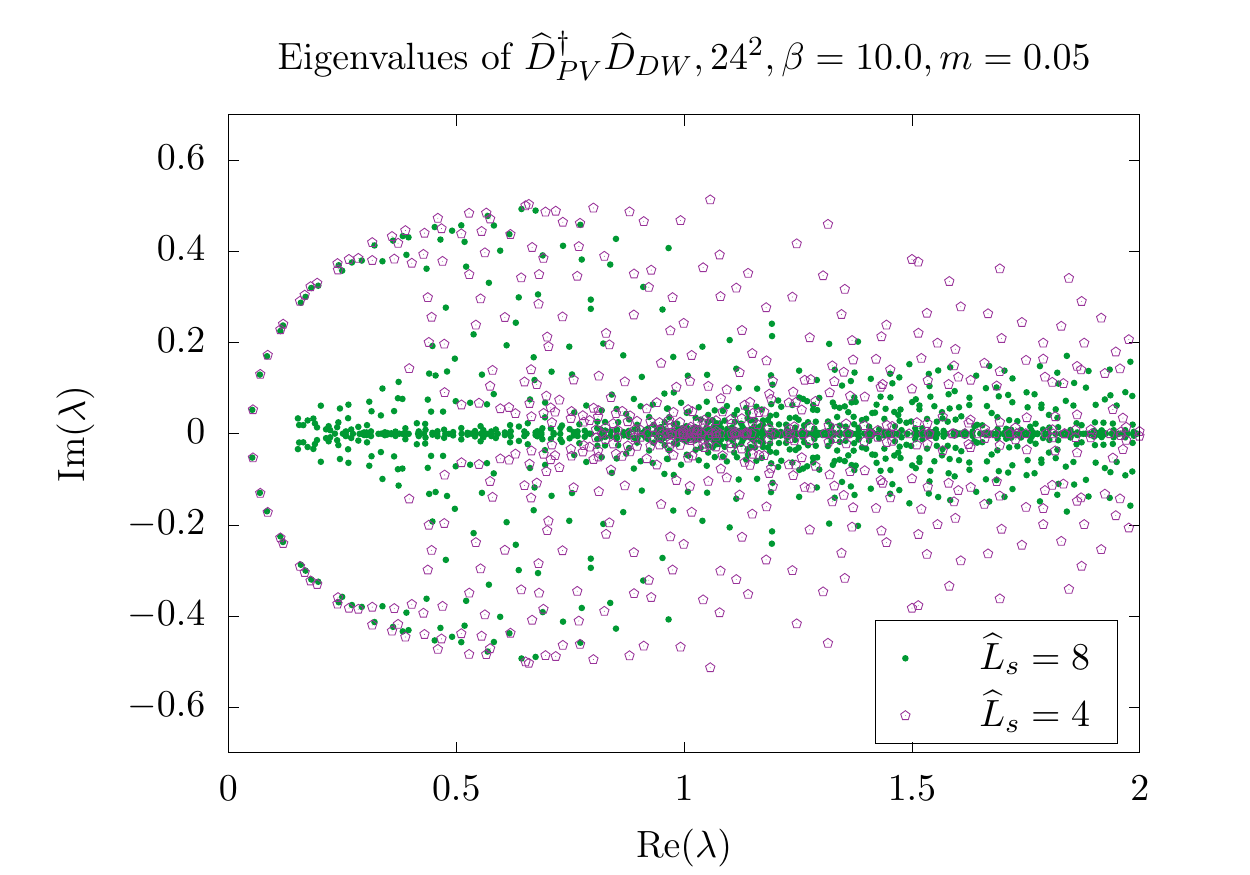}
\caption{On the left, the effect of the oblique projector for a boundary-only transfer operator for two value of the coarse $\widehat{L}_s$: the original 8, and a reduced 4. On the right, a comparison of the coarse spectrum for each of these choices of $\widehat{L}_s$.}
\label{fig:boundaryreduced}
\end{figure}

\subsection{Reduction of Coarse level $L_s$}
\label{sec:redcoarse}

An additional benefit of using a transfer operator which only acts on the boundaries is it gives us the flexibility to tune $L_s$ between levels. We will investigate this by two avenues. First, we will consider the behavior of the local colinearity and the oblique objector for two different values of the coarse $L_{s}$. Next, we will perform an explicit test of domain wall MG for a large value of the outer $L_s$, with a range of fixed smaller $L_s$ values on
all coarser levels.

\begin{figure}[ht] \centering
  \includegraphics[width=0.8\linewidth]{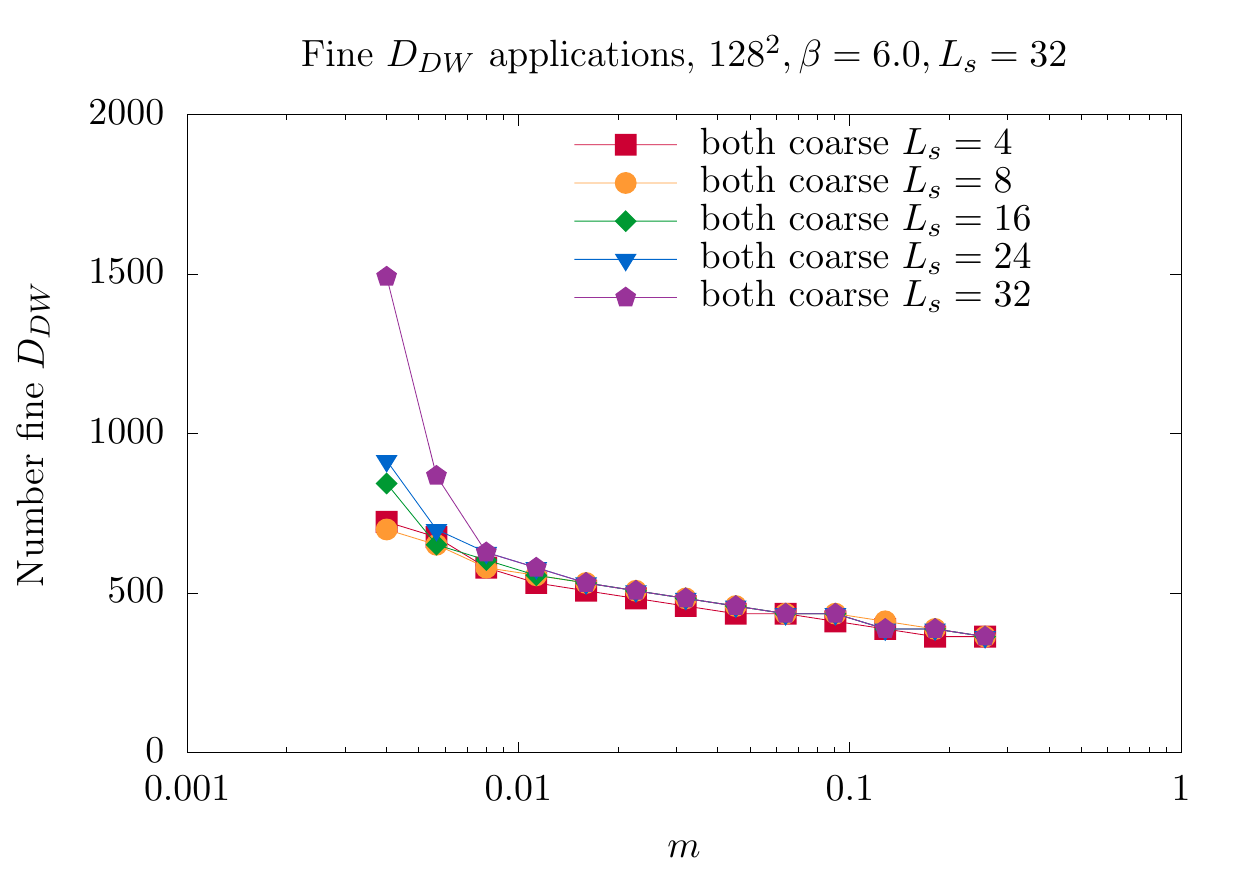}
  \caption{The number of applications of the fine domain wall operator ($D_{DW}$ and $D_{PV}^\dagger$ each count as one) per
    MG-preconditioned solve as a function of mass for fixed $\beta = 6.0$ and volume $128^2$. We vary the intermediate and coarse $L_{s}$ from 4 to 32. $M_5$ remains fixed.}
\label{fig:iterationsfinevaryls}
\end{figure}

We illustrate the oblique projector on the left-hand side of Fig.~\ref{fig:boundaryreduced}. We see that a reduced coarse $\widehat{L}_s = 4$ behaves at least as well as maintaining a constant $\widehat{L}_s = 8$. In addition, for larger values of the fine eigenvalue (i.e., higher momentum and bulk modes), the error enhancement is further suppressed. This may be related to the spectrum of the reduced $\widehat{L}_s$ operator on the right-hand of Fig.~\ref{fig:boundaryreduced}. We see that this operator does suffer from fewer spurious small eigenvalues, leading to a reduced risk of error enhancement. Following this investigation of the oblique projector and the spectrum on a small configuration, we have studied the performance of a solve on a $128^2$ configuration with a representative $\beta = 6.0$. Unlike our previous studies, we have chosen a large $L_{s} = 32$ for the outermost solve. For simplicity we chose a fixed reduced $L_{s}$ for both the intermediate and coarsest level.

We see in Fig.~\ref{fig:iterationsfinevaryls} that reducing the $L_{s}$ between the fine and the intermediate level \textbf{leads to a perfectly well behaved preconditioner}. This is impressive for two reasons. One, a reduction in $L_{s}$ does lead to an enhancement in chiral symmetry breaking, suggesting that the intermediate and coarsest levels would not accurately capture the low modes of the fine level. This does not appear to be a problem. Two, a reduced $L_{s}$ leads to enhanced stability for very small masses. There also  may be a side benefit of fewer spurious small modes for the coarsest operator with reduced $L_{s}$.

We are encouraged that reducing $L_{s}$ on the intermediate and coarsest level leads to an improved algorithm, and it is what informed the formulation studied in Sec.~\ref{sec:Results}. Of course, there is a wider parameter space we could explore in this study: varying the length of the extra dimension separately for each level, tuning $M_5$, tuning $m$ to approximately constant $m + m_{res}$, deflating the coarsest level to avoid a large iteration count. In light of our success simply reducing $L_s$ and making no other changes, we defer such in depth investigations to a study in four dimensions.

\newpage
\section{\label{sec:Conclusion}Conclusion}

We have presented a new approach to formulating an MG solver for domain wall fermions. This
is a critical step to realizing the full benefit of this chiral formulation which is theoretically 
superior but more computationally demanding than the Wilson and staggered discretizations. For clarity of exposition our formalism was restricted to the Shamir version of domain wall and for easy of development and  testing restricted to domain wall fermions for the \ndim{2} two-flavor Schwinger model. Neither of these choices are fundamental. These results convince us that this is a 
solution to a long sought fully recursive MG algorithm for domain wall fermions that can eliminate critical slowing down approaching the continuum limit for small fermion masses. 

As was the case with prototyping MG solvers the for Wilson~\cite{Brannick:2007ue}  and staggered~\cite{Brower:2018ymy} discretizations, the next step is to develop software and optimize performance for the Dirac solver for lattice QCD. We anticipate a range of algorithmic embellishments and software methods to optimize such an algorithm for use on exascale-trajectory machines in both the weak- and strong-scaling regimes. 

One salient feature is important to emphasize. The projection and prolongation only requires
finding the near null space of the 4-d Wilson kernel, saving computational overhead and memory occupancy relative to the na\"ive cost of the 5-d domain wall operator. The fact that there is no expansion of the null
space relative to Wilson MG due to the {\em heavy flavors} in the extra dimension is very good news.  We foresee that this approach will be effective even for workflows that require few solves per gauge field, e.g., Hybrid Monte Carlo.

This points to another benefit. The basic method presented here for the Shamir implementation applies equally well to M\"obius, Zolotarev, Borici, etc. formalism via the domain wall/Pauli-Villars factorization in~\BNO. The only requirement is that each factor is a linear functional of the Wilson kernel. As a consequence it should also be straight forward to generalize our algorithm directly to the {\bf overlap operator} itself. To appreciate the full landscape consider the large class of chiral fermion methods presented by Edwards, Joo, Kennedy, Orginos, and Wenger in Ref.~\cite{Edwards:2005an}.  In all of these the sign function $\mbox{\tt sign}[H]$ 
in the overlap operator
\be
 D_{ov} = \frac{1 + m}{2}  + \frac{1 - m}{2} \gamma_5 \mbox{\tt sign}[H]
\ee
 must be  approximated in a  variety of
 ways  as functional  of the Wilson operator: $\mbox{\tt sign}[H] \simeq 
 \epsilon[D_W]$ using, for example, rational approximations such as Pad\`{e} expansions, partial fractions, continued fractions, etc. The coarsening step would then act by projecting the \ndim{4} Wilson kernel into a near null space {\em before} building the  $\mbox{\tt sign}[H]$ function, 
 \be 
 \epsilon[D_W] \rightarrow  \widehat{\epsilon[D_W]} = \epsilon[\Pmg^\dag D_W \Pmg].
 \ee
As with Eq.~\ref{eq:KernelProject}, the kernel projection does {\bf not} commute with the effective chiral operator: $\epsilon[\Pmg^\dag D_W \Pmg]  \ne \Pmg^\dag\epsilon[ D_W ]\Pmg $. From this vantage point, our domain wall MG algorithm implementation is a special case using the domain wall/Pauli-Villars factorization in Eq.~\ref{eq:coarsenDW} to identify an effective overlap operator.
  
Clearly there is a larger landscape of MG algorithms for chiral fermion operators to explore. Optimizations will depend on the specific applications and target architectures. We anticipate that  alternative implementations of the MG solver for domain wall fermions in four dimensions~\cite{Cohen:2012sh,Boyle:2014rwa} may contribute to further optimizations. We leave the detailed study of these generalizations and optimizations for MG for domain wall and overlap chiral fermions to future investigations.
 
\section*{Acknowledgements}
We are grateful to Robert Edwards, Balint Joo, Harmut Neff, Kostas Orginos, and Pavlos Vranas for fruitful discussions. This work was supported in part by the U.S. Department
of Energy (DOE) under Award No. DE-SC0015845 and by the Exascale Computing Project (17-SC-20-SC), a collaborative effort of the U.S. Department of Energy Office of Science and the National Nuclear Security Administration.

\bibliographystyle{unsrt}
\bibliography{bib/MG,bib/moebius}

\begin{thebibliography}{10}

\bibitem{Brower:2018ymy}
Richard~C. Brower, M.~A. Clark, Alexei Strelchenko, and Evan Weinberg.
\newblock {Multigrid algorithm for staggered lattice fermions}.
\newblock {\em Phys. Rev.}, D97(11):114513, 2018.

\bibitem{Wilson:1974sk}
Kenneth~G. Wilson.
\newblock {Confinement of Quarks}.
\newblock {\em Phys.Rev.}, D10:2445--2459, 1974.

\bibitem{Blum:2013mhx}
T.~Blum, R.S. Van~de Water, D.~Holmgren, R.~Brower, S.~Catterall, et~al.
\newblock {Working Group Report: Lattice Field Theory}.
\newblock 2013.

\bibitem{Brower:1990at}
Richard~C. Brower, Claudio Rebbi, and Ettore Vicari.
\newblock {Projective multigrid for propagators in lattice gauge theory}.
\newblock {\em Phys. Rev. Lett.}, 66:1263--1266, 1991.

\bibitem{Wilson:1974mb}
Kenneth~G. Wilson.
\newblock {The Renormalization Group: Critical Phenomena and the Kondo
  Problem}.
\newblock {\em Rev. Mod. Phys.}, 47:773, 1975.

\bibitem{Brower:1991xv}
Richard~C. Brower, Robert~G. Edwards, Claudio Rebbi, and Ettore Vicari.
\newblock {Projective multigrid for Wilson fermions}.
\newblock {\em Nucl. Phys.}, B366:689--705, 1991.

\bibitem{Hulsebos:1990er}
Arjan Hulsebos, Jan Smit, and Jeroen~C. Vink.
\newblock {Multigrid inversion of the staggered fermion matrix}.
\newblock {\em Nucl. Phys. Proc. Suppl.}, 20:94--97, 1991.

\bibitem{Brannick:2007ue}
James Brannick, Richard~C. Brower, M~A. Clark, James~C. Osborn, and Claudio
  Rebbi.
\newblock {Adaptive Multigrid Algorithm for Lattice QCD}.
\newblock {\em Phys.Rev.Lett.}, 100:041601, 2008.

\bibitem{Babich:2010qb}
Ronald Babich, James Brannick, Richard~C. Brower, M~A. Clark, Thomas~A.
  Manteuffel, et~al.
\newblock {Adaptive multigrid algorithm for the lattice Wilson-Dirac operator}.
\newblock {\em Phys.Rev.Lett.}, 105:201602, 2010.

\bibitem{PhysRevD.11.395}
John Kogut and Leonard Susskind.
\newblock Hamiltonian formulation of wilson's lattice gauge theories.
\newblock {\em Phys. Rev. D}, 11:395--408, Jan 1975.

\bibitem{Kaplan:1992bt}
David~B. Kaplan.
\newblock {A Method for simulating chiral fermions on the lattice}.
\newblock {\em Phys.Lett.}, B288:342--347, 1992.

\bibitem{Cohen:2012sh}
Saul~D. Cohen, R.C. Brower, M.A. Clark, and J.C. Osborn.
\newblock {Multigrid Algorithms for Domain-Wall Fermions}.
\newblock {\em PoS}, LATTICE2011:030, 2011.

\bibitem{Boyle:2014rwa}
P~A Boyle.
\newblock {Hierarchically deflated conjugate gradient}.
\newblock 2014.

\bibitem{Yamaguchi:2016kop}
Azusa Yamaguchi and Peter Boyle.
\newblock {Hierarchically deflated conjugate residual}.
\newblock {\em PoS}, LATTICE2016:374, 2016.

\bibitem{Neuberger:1997bg}
Herbert Neuberger.
\newblock Vector like gauge theories with almost massless fermions on the
  lattice.
\newblock {\em Phys. Rev.}, D57:5417--5433, 1998.

\bibitem{Brannick:2014vda}
James Brannick, Andreas Frommer, Karsten Kahl, Björn Leder, Matthias Rottmann,
  and Artur Strebel.
\newblock {Multigrid Preconditioning for the Overlap Operator in Lattice QCD}.
\newblock {\em Numer. Math.}, 132(3):463--490, 2016.

\bibitem{Becher1982}
P.~Becher and H.~Joos.
\newblock The dirac-k{\"a}hler equation and fermions on the lattice.
\newblock {\em Zeitschrift f{\"u}r Physik C Particles and Fields},
  15(4):343--365, Dec 1982.

\bibitem{PhysRevD.38.1206}
Geoffrey~T. Bodwin and Eve~V. Kov\'acs.
\newblock Equivalence of dirac-k\"ahler and staggered lattice fermions in two
  dimensions.
\newblock {\em Phys. Rev. D}, 38:1206--1219, Aug 1988.

\bibitem{Brower:2012vk}
Richard~C. Brower, Harmut Neff, and Kostas Orginos.
\newblock {The Möbius domain wall fermion algorithm}.
\newblock {\em Comput. Phys. Commun.}, 220:1--19, 2017.

\bibitem{PhysRev.128.2425}
Julian Schwinger.
\newblock Gauge invariance and mass. ii.
\newblock {\em Phys. Rev.}, 128:2425--2429, Dec 1962.

\bibitem{Smilga:1996pi}
Andrei~V. Smilga.
\newblock {Critical amplitudes in two-dimensional theories}.
\newblock {\em Phys. Rev.}, D55:443--447, 1997.

\bibitem{Shamir:2000cf}
Yigal Shamir.
\newblock {New domain wall fermion actions}.
\newblock {\em Phys.Rev.}, D62:054513, 2000.

\bibitem{Brower:2004xi}
Richard~C. Brower, Hartmut Neff, and Kostas Orginos.
\newblock {Mobius fermions: Improved domain wall chiral fermions}.
\newblock {\em Nucl.Phys.Proc.Suppl.}, 140:686--688, 2005.

\bibitem{Brower:2005qw}
R.C. Brower, H.~Neff, and K.~Orginos.
\newblock {Mobius fermions}.
\newblock {\em Nucl.Phys.Proc.Suppl.}, 153:191--198, 2006.

\bibitem{Borici:1999da}
Artan Borici.
\newblock {Truncated overlap fermions: The Link between overlap and domain wall
  fermions {(\it in Lattice fermions and structure of the vaccuum, V. K.
  Mitrjushkin and G. Schierholz (eds))}}.
\newblock pages 41--52, 1999.

\bibitem{Borici:1999zw}
A.~Borici.
\newblock {Truncated overlap fermions}.
\newblock {\em Nucl.Phys.Proc.Suppl.}, 83:771--773, 2000.

\bibitem{Chiu:2002ir}
Ting-Wai Chiu.
\newblock Optimal domain-wall fermions.
\newblock {\em Phys. Rev. Lett.}, 90:071601, 2003.

\bibitem{Chiu:2002kj}
Ting-Wai Chiu.
\newblock Locality of optimal lattice domain-wall fermions.
\newblock {\em Phys. Lett.}, B552:97--100, 2003.

\bibitem{Edwards:2000rk}
Robert~G. Edwards and Urs~M. Heller.
\newblock Exact chiral symmetry for domain wall fermions with finite l(s).
\newblock {\em Nucl. Phys. Proc. Suppl.}, 94:737--740, 2001.

\bibitem{Edwards:2000qv}
Robert~G. Edwards and Urs~M. Heller.
\newblock Domain wall fermions with exact chiral symmetry.
\newblock {\em Phys. Rev.}, D63:094505, 2001.

\bibitem{Edwards:2005an}
Robert~G. Edwards, Balint Joo, Anthony~D. Kennedy, Kostas Orginos, and Urs
  Wenger.
\newblock {Comparison of chiral fermion methods}.
\newblock {\em PoS}, LAT2005:146, 2006.

\bibitem{Neuberger:1997fp}
Herbert Neuberger.
\newblock {Exactly massless quarks on the lattice}.
\newblock {\em Phys.Lett.}, B417:141--144, 1998.

\bibitem{Kikukawa:2000ac}
Yoshio Kikukawa and Tatsuya Noguchi.
\newblock {Low-energy effective action of domain wall fermion and the
  Ginsparg-Wilson relation}.
\newblock {\em Nucl. Phys. B Proc. Suppl.}, 83:630--632, 2000.

\bibitem{Ginsparg:1981bj}
Paul~H. Ginsparg and Kenneth~G. Wilson.
\newblock A remnant of chiral symmetry on the lattice.
\newblock {\em Phys. Rev.}, D25:2649, 1982.

\bibitem{doi:10.1002/gamm.201490008}
Jörg Liesen and Petr Tichý.
\newblock Convergence analysis of krylov subspace methods.
\newblock {\em GAMM-Mitteilungen}, 27(2):153--173, 2004.

\bibitem{benson1973iterative}
M.W. Benson.
\newblock {\em Iterative Solution of Large Scale Linear Systems}.
\newblock Mathematics report. Thesis (M.Sc.)--Lakehead University, 1973.

\bibitem{Chow:2001:PIP:1080623.1080641}
Edmond Chow.
\newblock Parallel implementation and practical use of sparse approximate
  inverse preconditioners with a priori sparsity patterns.
\newblock {\em Int. J. High Perform. Comput. Appl.}, 15(1):56--74, February
  2001.

\bibitem{Sterck06reducingcomplexity}
Hans~De Sterck, Ulrike~Meier Yang, Jeffrey, and J.~Heys.
\newblock Reducing complexity in parallel algebraic multigrid preconditioners.
\newblock {\em SIAM J. Matrix Anal. Appl}, 27:1019--1039, 2006.

\bibitem{doi:10.1137/140952570}
Eran Treister and Irad Yavneh.
\newblock Non-galerkin multigrid based on sparsified smoothed aggregation.
\newblock {\em SIAM Journal on Scientific Computing}, 37(1):A30--A54, 2015.

\bibitem{Kaplan:1999jn}
David~B. Kaplan and Martin Schmaltz.
\newblock {Supersymmetric Yang-Mills theories from domain wall fermions}.
\newblock {\em Chin.J.Phys.}, 38:543--550, 2000.

\bibitem{Callan:1984sa}
Curtis~G. Callan, Jr. and Jeffrey~A. Harvey.
\newblock {Anomalies and Fermion Zero Modes on Strings and Domain Walls}.
\newblock {\em Nucl. Phys.}, B250:427--436, 1985.

\bibitem{golub13}
Gene~H. Golub and Charles~F. van Loan.
\newblock {\em Matrix Computations}.
\newblock JHU Press, fourth edition, 2013.

\bibitem{Adams:2009eb}
David~H. Adams.
\newblock {Theoretical foundation for the Index Theorem on the lattice with
  staggered fermions}.
\newblock {\em Phys. Rev. Lett.}, 104:141602, 2010.

\bibitem{Vranas:1997da}
Pavlos~M. Vranas.
\newblock Chiral symmetry restoration in the schwinger model with domain wall
  fermions.
\newblock {\em Phys. Rev.}, D57:1415--1432, 1998.

\end{thebibliography}

\appendix
\section{Low Momentum Expansions for the DW Fermions}
\label{app:lowepequiv}

The free theory, setting $U = 1$, can be expanded and diagonalized in momentum space for finite $L_s$, giving  valuable guidance to our MG construction. First consider our approximation, 
 $D^\dag_{PV} \simeq  D^{-1}_{PV} $, which we will  prove  to valid up to error $\mathcal{O}(p^4)
$ at finite  finite $L_s$.  Indeed the Pauli-Villars operator is especially
simple because it is anti-periodic in $L_s$, and thus even with interacting fields can be diagonalized via Fourier modes giving
\be
D_{PV}(U, p_5) = \gamma_5 \sin(p_5) + 1 - \cos(p_5)  +  M_5+ D_W(U,0),
\ee
in terms of $d \times d$ space-time blocks for the Wilson operator $D_W(U,0)$.  Setting $M_5 = -1$,
\be
D_{PV}(U, p_5) = - e^{-i p_5 \gamma_5} + D_W(U,0),
\ee
we note that the first term by in isolation gives a circle of eigenvalues in the complex plane. Moreover, as is apparent in Fig.~\ref{fig:dw_spectra}, this basic pattern persists even with non-trivial gauge fields. In the free-field limit, we can further diagonalize in space-time Fourier modes, $p_\mu$, giving
\be
\widetilde D_{PV}(p_\mu,p_5)  =   i \sum^5_{M=1}\gamma_M \sin( p_M) +  \sum^5_{M=1} 2 \sin^2( p_M/2) + M_5,
\ee
where the summation includes $\gamma_5$ and $p_5$. The free normal operator is
\be
\widetilde D_{PV}^\dagger \widetilde D_{PV} =   \sum^5_{M=1}\sin^2( p_M) + (  \sum^5_{M=1} 2 \sin^2( p_M/2) + M_5)^2 \; , 
\ee
diagonal in spin structure. In the case of $M_5 = -1$, the low momentum expansion of this operator is $\widetilde D_{PV}^\dagger \widetilde D_{PV} = M^2_5 + \mathcal{O}(p_M^4)$. This gives
\be
\widetilde D^\dag_{PV} =  \widetilde D^{-1}_{PV} + \mathcal{O}(p^4).
\ee 
Indeed more generally for $M_5 = -1$ one can prove
that $D_{PV}^\dagger \widetilde D_{PV}$ is bounded below by 1, i.e., at the lattice cutoff scale $1/a^2$. With 
non-trivial gauge fields the approximation $\widetilde D^\dag_{PV} \approx  \widetilde D^{-1}_{PV}$ holds qualitatively particularly when $M_5$ is
appropriately tuned.

The next step is to compare the low eigenspectra of the
preconditioned operator $D^{-1}_{PV} D_{DW}(m)$ and the approximation
$D^\dag_{PV} D_{DW}(m)$. From Fig.~\ref{fig:overlap_and_approx} we note
the near exact coincidence of small eigenvalues even in the
presence of gauge fields. We can elucidate this in the free-field
limit.

We cannot simultaneously diagonalize $D_{PV}$ and $D_{DW}(m)$ in the bulk dimension due to the difference in boundary conditions which complicates the analysis. In this case we take advantage of low-momentum perturbation theory in the space-time dimensions. Given $\widetilde D_{DW}(p_\mu, m)_{ss'}$, we expand both it and the Pauli-Villars factor at low d-momenta, take the appropriate products and solve the
characteristic polynomial for the eigenvalues. The pairing of low modes gives a complex square root singularity in the complex plane, leading to the circular structure of the low spectrum as evident in Fig.~\ref{fig:overlap_and_approx}.  The results from this
approach are given in the text in Eq.~\ref{eq:ExactLowEV} and 
Eq.~\ref{eq:ApproxLowEV}.  Comparing the exact vs the approximate low
spectra we see they are identical up to a six-dimensional operator
$\mathcal{O}(a^2 m p^2)$.

We additionally note that the low momentum expansion of the effective overlap operator,
\be
D_{ov}(m) =\frac{1 + m }{2}+ \frac{1 -m }{2} \gamma_5 \epsilon_L[H]
\quad, \quad \epsilon_L[H] = \frac{(1 - H)^L - (1 + H)^L}{(1 - H)^L +  (1 + H)^L} \; ,
\label{eq:Overlap}
\ee
has universal coefficients even at finite $L_s$ up to $\mathcal{O}(p^n)$ for $L_s > n$. As a result, the eigenvalues are also equivalent in perturbation theory up to that order. For example,
consider for $d=2$ the roots of the polynomial, 
$\lambda^2  -  \lambda \;\mbox{tr}  +  \mbox{det}= 0$ in terms
of the trace and determinant. For the Shamir kernel, Evaluating this for the Shamir kernel, $ H = \gamma_5 D_W(M_5)/( 2 + D_W(M_5))$, this is given by
\be
\mbox{det} =\lambda_+ \lambda_- = p_x^2 + p_y^2 - (4/3)[ (p_x^2 + p^2_y)^2  -
p_x^2 p_y^2 ]\quad, \quad  \mbox{tr} =   \lambda_+ + \lambda_- = 2 (p_x^2 + p_y^2)  \; 
\ee
up to $\mathcal{O}(p^2)$, again universal for any $L_s \ge 2$, resulting in the eigenvalues 
\be
\lambda_{\pm} =   \pm i \sqrt{ p^2(1 - p^2) + 2p^2_x p^2_y/3 }
+  p^2 \;,
\ee 
to the same order. To include non-zero mass
we can use the mass shift identity for
eigenvalus, $\lambda^0 \rightarrow \lambda = m + (1-m)\lambda^0$,  from  Eq.~\ref{eq:DWoperator} to get 
\be 
\lambda_{\pm} = m +p^2 \pm i (1-m)\sqrt{ p^2(1 - p^2) +
  2p^2_x p^2_y/3 } - m p^2   \; ,
\ee  
which up to  rescaling by
$1/(1 -m)$ gives $\lambda_{\pm} \sim  m_q  +  p^2 \pm i\sqrt{ p^2} $
in agreement with Eq.~\ref{eq:ExactLowEV} in the text up to a dimension-three scaling operator of  $\mathcal{O}(m p^2)$.
Higher-order expansions in 2-d and 4-d effective overlap operators
are easily shown in Mathematica  to follow this expansion
invariance up to $\mathcal{O}(p^{L_s}) $. We believe the fact
that the quadratic form is independent  of  $L_s \ge 2$ even for 
our approximate $D^\dag_{PV} D_{DW}(m)$ effective overlap
operator is at the core the effectiveness of using small $L_s$ for
MG iterations on the coarse level.

Finally we note that this expansion is almost certainly a divergent
asymptotic series and as such does not by itself lend itself to
fix the coefficients in improved higher-order polynomial approximation
\be
[D_{PV} D^\dag_{PV}]^{-1} \simeq c_0 +c_1 D_{PV} D^\dag_{PV} + c_2 (D_{PV} D^\dag_{PV})^2 \cdots
\ee
to
$[D_{PV} D^\dag_{PV}]^{-1}$ in Eq.~\ref{eq:NormalExp} beyond the zeroth order. We see this first-order equivalence is given by $c_i= (1,0,0,0\cdots)$. We can take thus further by expanding $[ D_{PV} D^\dag_{PV}]^{-1} = 1/( 1 + A _{PV})$ in
$A _{PV} = D_{PV} D^\dag_{PV} -1$ to show that
each term in Eq.~\ref{eq:NormalExp} corrects
the approximation by  another two powers of $p$. For
example just the second-order expansion,  $c_i= (2,-1,0,0,\cdots)$, giving
the equivalence $ D^{-1} _{PV}  =   D_{PV}^\dag(2  - D_{PV}
D^\dag_{PV}) + \mathcal{O}(p_\mu^4) $. This polynomial
on its own cannot be used in our MG algorithm because $2  - D_{PV}
D^\dag_{PV}$ is not positive definite so coefficient in a polynomial truncation  must be
weighted appropriately over the entire spectrum. 

\section{Proof that the spectrum has a positive real part}
\label{app:RHPproof}

In the full interacting case with gauge fields we consider our
approximation  $D_{PV}^\dagger D_{DW}$ as $\left(D_{PV}^\dagger D_{PV}\right) D_{PV}^{-1} D_{DW}$ and note from Eq.~\ref{eq:DWFmatrix} that $D_{PV}^{-1} D_{DW}$ has the eigenspectra of the finite-$L_s$ overlap kernel,
\be
D_{ov}^{(L_s)}(m) = \frac{1+m}{2} + \frac{1-m}{2} \gamma_5 \epsilon_{L_s}\left[H_5\right] \; ,
\ee
plus additional $1$ eigenvalues. We recall that for even $L_s$, $\epsilon_{L_s}\left[H_5\right] \in [-1,1]$.  Applying the Cauchy-Schwarz  $\gamma_5 \epsilon_{L_s}\left[H_5\right]$ inequality  for a normalized vectors, $\<u|u\> = \<v|v\> =1$, the matrix element is bounded in magnitude,
\be
|\bra{u} \gamma_5 \epsilon_{L_s}\left[H_5\right]\ket{v}|= \sqrt{\braket{v|\epsilon_{L_s}\left[H_5\right]\epsilon_{L_s}\left[H_5\right]|v}}\le 1 \; ,
\ee
Equivalently this is just a  statement of the unitarity bound on matrix elements. 
By extension, the spectrum of $D_{ov}(m)$ lives inside a circle of radius $\frac{1-m}{2}$ centered at $(\frac{1+m}{2},0)$ in the complex plane. Thus
both $D_{ov}(m)$ and $D_{PV}^{-1} D_{DW}$ have  positive definite real part for $m>0$. 

Let us now consider the eigenvalue problem for the operator $D_{PV}^\dagger D_{DW}$ for {\emph{any}} right eigenvector $\ket{\lambda}$,
\be
 D_{PV}^\dag D_{DW} \ket{\lambda} = r e^{i\theta} \ket{\lambda}\; .
\ee
where we have generally written the eigenvalue $\lambda$ as $r e^{i \theta}$. We wish to prove the right-half plane condition $\theta \in (-\pi/2, \pi/2)$. We note we can re-write the left-hand side of the above system as
\be
 \left(D_{PV}^\dagger D_{PV}\right) D_{PV}^{-1} D_{DW \ket{\lambda}} = r e^{i\theta} \ket{\lambda}\; .
\ee
Left multiplying by $\left(D_{PV}^\dagger D_{PV}\right)^{-1}$ and taking the full matrix element with $\bra{\lambda}$ we have 
\be
\bra{\lambda} D_{PV}^{-1} D_{DW} \ket{\lambda} = r e^{i \theta} \bra{\lambda} (D_{PV}^\dagger D_{PV})^{-1} \ket{\lambda}.
\ee
Since $D_{PV}^{-1} D_{DW}$ satisfies the right half-plane condition and $D_{PV}^\dagger D_{PV}$ is a Hermitian positive operator, this proves the right half-plane condition  $\theta \in (-\pi/2,\pi/2)$ for $m > 0$.

\end{document}